\newcommand{\cB}[0]{{\mathcal B}}
\newcommand{\cH}[0]{{\mathcal H}}
\newcommand{\cJ}[0]{{\mathcal J}}
\newcommand{\cK}[0]{{\mathcal K}}
\newcommand{\cL}[0]{{\mathcal L}}
\newcommand{\cM}[0]{{\mathcal M}}

\newcommand{\lC}[0]{{\mathbb C}}
\newcommand{\lL}[0]{{\mathbb L}}

\newcommand{\ket}[1]{\left|\, #1\right\rangle}
\newcommand{\bra}[1]{\left\langle #1\right|}
\newcommand{\bracket}[2]{\left.\left\langle #1\right|#2\right\rangle}
\newcommand{\Hom}[0]{{\rm Hom}\,}
\newcommand{\Tr}[0]{{\rm Tr}\,}

\newcommand{\essence}[0]{essence}
\newcommand{\physical}[0]{physical}
\newcommand{\virtual}[0]{virtual}
\newcommand{\conjugacy}[0]{conjugacy}

\newcommand{\hlu}[2]{_{#1}{}^{#2}}
\newcommand{\cotimes}[0]{\otimes\cdots\otimes}
\newcommand{\db}[0]{\(d\)-bein}

\documentclass{article}
 \usepackage[hypertex]{hyperref}
\usepackage{amssymb,graphicx,amsmath}
\title{The props of quantum mechanics}
\author{George Svetlichny\footnote{Departamento de Matem\'atica, Pontif\'{\i}cia Universidade Cat\'olica, Rio de Janeiro, Brazil \newline
svetlich@mat.puc-rio.br \hfill \url{http://www.mat.puc-rio.br/\~svetlich}}}
\begin{document}
\maketitle

\begin{abstract}
We introduce a formalism that exploits the many-input many-output nature of nodes in quantum circuits. There is a diagrammatic and an algebraic version, the latter similar to the spinor formalism of general relativity. This allows us to work in truly basis independent ways, clarifying and simplifying many aspects of quantum state processing. The narrative is at times interrupted by antics of characters from  quantum age fairy tales.
\end{abstract}
\section{Introduction}\label{intro}

The title of this article has triple meaning. Firstly, a {\em prop\/} is a mathematical structure abstracted from the compositional structure of many-valued many-variable functions, which in turn is a generalization of an {\em operad\/} the abstraction of a similar structure of single valued many-variable functions, this, in its turn, being a generalization of {\em categories\/} with which we assume the reader has some familiarity. Props are also theatrical objects, and we use  this term metaphorically for the physical objects and devices, such as lasers, crystals, measuring apparatus, etc.\ that have to be present on the laboratory ``stage'' for quantum mechanics to play its role. Finally, props are meant as the mental devices we lean upon to achieve some semblance of understanding of the play. Part of these are all the mathematica tools, and part  the various ``interpretations" of quantum mechanics, such as Copenhagen, many worlds, many minds, coherent histories, QBism, etc.\footnote{Probably as many variants as there are thinkers of things quantum.} In this paper we are primarily interested in the mathematics of quantum mechanics and so we'll adopt a radical version of what is affectionately know by some as the ``shut up and calculate" interpretation. We abbreviate this to {\sl SHUAC\/}\footnote{I like to pronounce it thus: {\sl Shoe-ack}. Footnotes will be written in the first person singular. This is one of the theatrical props of this play, i.e., article. The SHUAC mathematician will be our guide at various points in this play.} as from time to time we'll want to refer to it.

We shall not give much detail concerning the mathematical prop of quantum mechanics as much of this is still to be worked out. We shall work only with the concrete example of Hilbert spaces and maps between tensor products of such. A proper generalization would in principle extend the present categorical approaches.\footnote{For a list of references see the Wikipedia article ``Categorical quantum mechanics".} Though at times we do make category theoretical remarks, no knowledge of category theory is needed to understand all the main points of this paper. We shall need an adjective to correspond to {\em prop\/} and have adopted {\em propic\/}.\footnote{It rhymes with {\em tropic}, which adds to it's appeal. One could have used {\em propical\/} which rhymes with {\em tropical\/} but I did not want to be too categorical.} Our treatment of props is greatly oversimplified and readers familiar with them might feel we're not being fair neither to the concept nor the spirit. We're basically emphasizing the many-to-many nature of the object handled by prop theory, typically maps between finite tensor product of algebras. We are deliberately not introducing much structure, feeling that such a minimalist approach will more easily reveal what is truly intrinsic to quantum mechanics unencumbered by an overly formalized exposition.

We begin in Section \ref{FinHprop} by diagrams and notation.  The diagrammatic approach, reminiscent of Feynman diagrams, is used to express common situations arising in quantum information theory and is similar to other such practices in the literature. We also borrow a notation from the spinor formalism of general relativity. These tools are to a large extent basis independent in contrast to much of quantum information literature. This lends it greater power to reach the necessary conclusions. The same diagram or algebraic expression can lend itself to various alternative interpretations such as ``state'', ``channel'', ``amplitude'', etc.\ allowing for greater clarity and analytical power of treatment as exemplified in Section \ref{indep}. Much of quantum mechanical literature has a fairy-tale-like character\footnote{Some feel it's like science fiction, but fairy tale seems more appropriate. Alice and Bob would agree.} and in Section \ref{tale} we tell a tall tale about the creation of quantum teleportation. The characters of this fairy tale and other embedded stories will occasionally interrupt the main exposition to make us take note of what could have been missed otherwise. In Section \ref{whatswhat} we explore some ontological questions arising from the propic nature of tensor products of finite dimensiona Hilbert spaces, specifically what concerns causality, time, and locality. The prop approach sheds new light on these notions. In section \ref{woods} the plot thickens. We go up one step in the ladder \(\cH\to \cB(\cH) \to \cB(\cB(\cH))\to \cdots\) and look into time-travel.

\section{The prop of finite dimensional Hilbert spaces}\label{FinHprop}

We shall deal with finite dimensional complex Hilbert spaces, their tensor products and linear maps between such products. Each Hilbert space is either {\em \physical\/}, meaning that it represents something physically present in the laboratory, that is, corresponds to a laboratory prop, or else, {\em \virtual\/} when it is the dual of a \physical\ space. The dual of a \virtual\  space we shall take to be the corresponding \physical\ space.

By a tensor product we shall mean the tensor product of any finite number of Hilbert spaces, each of which may be either \physical\ or \virtual\ .  If \(P\) and \(Q\) are two such products we denote by \(\cL(P,Q)\) the set of linear maps between the two, and write \(\cL(P)\) for \(\cL(P,P)\). We shall now introduce a diagrammatic way of representing elements of such spaces, their relations and compositions. We'll call such diagrams {\em propic diagram\/}.
An element \(L\in \cL(P,Q)\) shall be represented diagrammatically by a simple closed curve or polygon with certain incoming and/or outgoing lines labeled by the individual spaces in the two products. We shall call such a closed curve or polygon a {\em node\/}. Arrows on the lines indicate if the space belongs to \(P\) or to \(Q\), those with incoming arrows belong to \(P\) and with outgoing belong to \(Q\). Solid lines are used for \physical\ spaces and dotted lines for \virtual\ . For instance an element \(L\in \cL(H_1\otimes H_2^*, H_1\otimes H_3\otimes H_4^*)\) would be represented by Figure \ref{node}:
\newsavebox{\anode}
\savebox{\anode}{
\put(50,50){\circle{40}}
\put(15,45){\vector(3,1){0}}\put(0,40){\line(3,1){30}}
\qbezier[20](10,10)(15,15)(35,35)\put(25,25){\vector(1,1){0}}
\qbezier[20](65,35)(75,25)(90,10)\put(75,25){\vector(1,-1){0}}
\put(35,65){\line(-1,1){25}}\put(25,75){\vector(-1,1){0}}
\put(15,40){\makebox(0,0){\small \(1\)}}
\put(17,10){\makebox(0,0){\small \(2\)}}
\put(33,75){\makebox(0,0){\small \(1\)}}
\put(85,22){\makebox(0,0){\small \(4\)}}
\put(50,50){\makebox(0,0){\(L\)}}
}
\begin{figure}[h!]
\begin{center}
\begin{picture}(100,100)(0,0)
\put(0,0){\usebox{\anode}}
\put(85,55){\makebox(0,0){\small \(3\)}}
\put(70,50){\line(1,0){30}}\put(85,50){\vector(1,0){0}}
\end{picture}
\end{center}
\caption{A node.\label{node}}
\end{figure}

For simplicity we have labeled each lines with the index of the corresponding Hilbert space. We shall forgo labels when the context supplies them.

Directions on the page (up, down, right, left, etc.) have no significance, nor the order of attachments to the border of the node. Appropriate canonical permutation equivalences among tensor factors are understood to apply when interpreting the diagrams. The same diagram above will thus also represent the image of \(L\) in \(\cL( H_2^*\otimes H_1, H_1\otimes H_3\otimes H_4^*)\) or \(\cL(H_1\otimes H_2^*,  H_3\otimes H_1\otimes H_4^*)\), etc.\ under canonical isomorphisms. This practice is ambiguous when two lines of the same \essence\ (\physical\ or \virtual) are attaches to a node and carry the same Hilbert space label. For instance, without additional notation one would not distinguish in \(\cL(\cH\otimes \cH)\) the identity map \(e\otimes f\mapsto e\otimes f\) from the exchange map \(e\otimes f\mapsto f\otimes e\). We shall not introduce any scheme to resolve this and other ambiguities so as not to overburden the diagrams and other notation. Proper explanations at appropriate times will prevent any misunderstandings.

Nodes with no incoming lines represents a map from the one-dimensional Hilbert space \(\lC\) to the product designated by the outgoing lines, and a node with no outgoing lines is a map from the product to \(\lC\). Since \(\lC\otimes \cH \simeq \cH \simeq \cH\otimes \lC\) the space \(\lC\) is generally not represented by anything. It is considered both a \physical\ space and a \virtual\  (the only one such) as there is a canonical {\em linear\/} duality between the two, given by the identity map. Each  \(\cL(P,Q)\) is itself a Hilbert space with the inner product given by
\begin{equation}\label{inner}
(A,B)=\Tr(A^*B)
\end{equation}
 for \(A,\,B\in \cL(P,Q)\) and \(A^*\) denoting the adjoint of \(A\).

We shall make systematic use of the basic defining mathematical property of tensor products, called {\em  universality\/}. We recall that if   \(V_1,\dots,V_n\) are vector spaces, their {\em  tensor product\/} is a vector space usually denoted by \(V_1\cotimes V_n\) along with an \(n\)-linear map \(J:V_1\times \cdots \times V_n\to V_1\cotimes V_n\) such that {\em  any\/} \(n\)-linear map \(\alpha:V_1\times \cdots \times V_n\to W\) to yet another vector space \(W\) factors {\em uniquely\/} though a {\em  linear\/} map \(\hat\alpha:V_1\otimes \cdots \otimes V_n\to W\), that is \(\alpha=\hat\alpha\circ J\). In other words \(J\) is a {\em  universal\/} \(n\)-linear map and any other differs from it by a {\em  unique\/} subsequent {\em  linear\/} factor. One generally writes \(v_1\cotimes v_n\) for \(J(v_1,\dots,v_n)\).
Recall also that a multipartite quantum state-vector resides in a tensor product Hilbert space \(\cH_1\cotimes\cH_n\) where each \(\cH_i\) is the Hilbert space of states of the \(i\)-th part. States of the form  \(\phi _1\otimes\phi _2\cotimes\phi _n\) are called {\em  product\/}, or {\em  disentangled\/} states while all states that cannot be put into this form are called {\em  entangled\/}.
Universality has at least two interesting consequences: (1) any linear construct on entangled states is uniquely determined by what it does on disentangled states; (2) any theorem that uses only linearity on entangled states is true if it is true on disentangled states. These facts can considerably simplify definitions, constructions, and proofs.
All of the above is also true if we systematically replace the word ``linear" by ``antilinear" (with \(J\) still \(n\)-{\em linear\/}).

We shall make a distinction between {\em naming\/} a function and {\em expressing\/} it. To name it is simply to designate it by a symbol, whereas to express it is to somehow designate it's action. Thus given \(f:X\to Y\), naming it is to simply write \(f\) whereas expressing it is to write \(f(\cdot)\) meaning that it receives an argument indicated by the dot. Naming the composition of a function \(g\) followd by \(f\) is to write \(f\circ g\) and expressing it is to write \(f(g(\cdot))\). If \(e\in\cH\) is an element of a Hilbert space, it's a Riez representative of a linear functional in \(\cH^*\) whose name we shall take to be \(\bar e\) and whose expression is \((e,\cdot)\).

One has the canonical isomorphism:
\begin{equation}\label{caniso}
\cL(\cH_1\otimes K,\cH_2)\cong \cL(\cH_1,\cH_2\otimes K^*).
\end{equation}
Any element of the space on the left-hand side is a linear sum of elements of the form \((h_1,\cdot)(k,\cdot)h_2\). This is an {\em expressed\/} function meaning it maps \(e\otimes f\in \cH_1\otimes \cK\) to \((h_1,e)(k,f)h_2\). The above canonical isomorphism is then given by:
\begin{equation}\label{eq:switcheroo}
(h_1,\cdot)(k,\cdot)h_2\mapsto (h_1,\cdot)h_2\otimes \bar k,
\end{equation}
which is nothing more than switching the expressed function \((k,\cdot)\) to its name (and introducing an appropriate tensor product symbol). By tensor universality, (\ref{eq:switcheroo}) defines a linear map uniquely. It is obviously bijective as the reverse switch is just as well defined.

What this isomorphism means is that in relation to a node, one can change  any incoming or any outgoing line to it opposite direction, provided we also change the attribute of being \physical\ or \virtual\  to it's opposite.\footnotemark\  By the {\em direction\/} of a line we shall mean it's attribute of being incoming or outgoing, and by its {\em \essence\/} it's attribute of being \physical\ or \virtual. An {\em opposite\/} line is then one with both attributes changed.

\begin{figure}[h!]
\begin{center}
\begin{picture}(100,100)(0,0)
\put(50,50){\circle{40}}
\qbezier[20](0,40)(15,45)(30,50)\put(15,45){\vector(-3,-1){0}}
\put(10,10){\line(1,1){26}}\put(23,23){\vector(-1,-1){0}}
\qbezier[20](65,35)(75,25)(90,10)\put(75,25){\vector(1,-1){0}}
\put(35,65){\line(-1,1){25}}\put(25,75){\vector(-1,1){0}}
\put(70,50){\line(1,0){30}}\put(85,50){\vector(1,0){0}}
\put(15,40){\makebox(0,0){\small \(1\)}}
\put(17,10){\makebox(0,0){\small \(2\)}}
\put(33,75){\makebox(0,0){\small \(1\)}}
\put(85,55){\makebox(0,0){\small \(3\)}}
\put(85,22){\makebox(0,0){\small \(4\)}}
\put(50,50){\makebox(0,0){\(L\)}}
\end{picture}
\hskip 60pt
\begin{picture}(100,100)(0,0)
\put(50,50){\circle{40}}
\put(15,45){\vector(3,1){0}}\put(0,40){\line(3,1){30}}
\put(10,10){\line(1,1){26}}\put(23,23){\vector(-1,-1){0}}
\put(65,35){\line(1,-1){25}}\put(75,25){\vector(-1,1){0}}
\put(35,65){\line(-1,1){25}}\put(25,75){\vector(-1,1){0}}
\put(70,50){\line(1,0){30}}\put(85,50){\vector(1,0){0}}
\put(15,40){\makebox(0,0){\small \(1\)}}
\put(17,10){\makebox(0,0){\small \(2\)}}
\put(33,75){\makebox(0,0){\small \(1\)}}
\put(85,55){\makebox(0,0){\small \(3\)}}
\put(85,22){\makebox(0,0){\small \(4\)}}
\put(50,50){\makebox(0,0){\(L\)}}
\end{picture}
\end{center}
\caption{Two other versions of \(L\) of Fig. \ref{node}. \label{nodeversions}}
\end{figure}
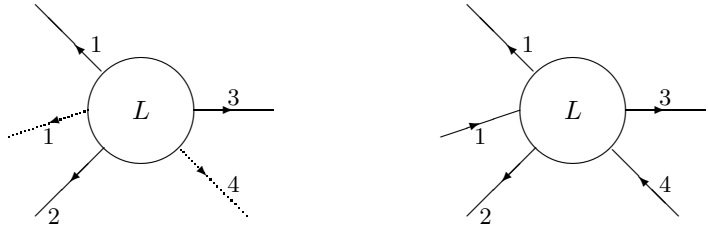

When the attributes of the lines of a node are changed in the above way, we shall continue to designate it by the same letter. This will not cause any confusion or ambiguity if a few simple rules are followed. Two other versions of the node of Figure \ref{node} are shown in Fig. \ref{nodeversions}. In the first of these we've made all the lines
outgoing, and in the second we've made all the lines \physical.

The first one now represents a ``state" in \(\cH_1\otimes\cH_1^*\otimes \cH_2\otimes\cH_3\otimes\cH_4^*\) and we wrote ``state" in quotes as some of the Hilbert spaces in this product are \virtual. The second version can be construed as a channel from \(\cH_1\otimes\cH_4\) to \(\cH_1\otimes \cH_2\otimes\cH_3\). The various ideas of equivalence of states and channels that one meets in quantum information theory are all consequences of  (\ref{caniso}).

If an outgoing line of one node carries the same Hilbert space label as the incoming one of another, and the two lines have the same essence, then the nodes can be joined by joining the two lines to form a composite node. Thus say we have the node in Fig. \ref{anothernode},
\newsavebox{\anothernode}
\savebox{\anothernode}{
\put(50,50){\circle{40}}
\qbezier[20](65,35)(75,25)(90,10)\put(75,25){\vector(-1,1){0}}
\put(36,64){\line(-1,1){25}}\put(25,75){\vector(-1,1){0}}
\put(33,75){\makebox(0,0){\small \(2\)}}
\put(85,22){\makebox(0,0){\small \(5\)}}
\put(50,50){\makebox(0,0){\(M\)}}
}
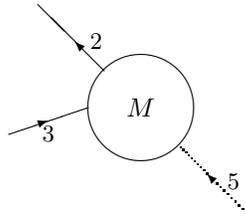
\begin{figure}[h!]
\begin{center}
\begin{picture}(100,100)(0,0)
\put(0,0){\usebox{\anothernode}}
\put(15,45){\vector(3,1){0}}\put(0,40){\line(3,1){30}}
\put(15,40){\makebox(0,0){\small \(3\)}}
\end{picture}
\end{center}
\caption{Another node.\label{anothernode}}
\end{figure}

\noindent then this node can be joined to that of Fig. \ref{node} to obtain the composite given in Fig. \ref{nodecomposition}.
\noindent The relative position of the two nodes in the diagram has no
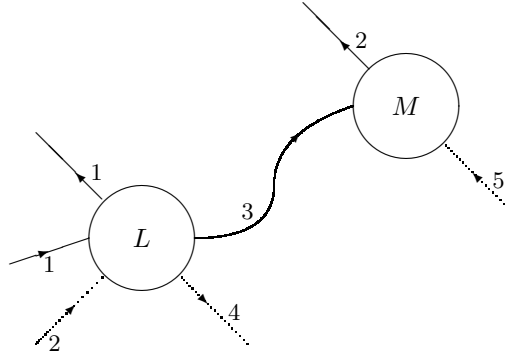
\begin{figure}[h!]
\begin{center}
\begin{picture}(200,150)(0,0)
\put(0,0){\usebox{\anode}}
\put(100,50){\usebox{\anothernode}}
\qbezier(70,50)(100,50)(100,70)\qbezier(100,70)(100,90)(130,100)
\put(110,90){\vector(1,1){0}}
\put(90,60){\makebox(0,0){\small \(3\)}}
\end{picture}
\end{center}
\caption{Composite node.\label{nodecomposition}}
\end{figure}

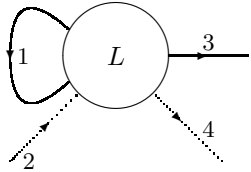
\begin{figure}[h!]
\begin{center}
\begin{picture}(100,100)(0,0)
\put(50,50){\circle{40}}
\qbezier[20](10,10)(15,15)(35,35)\put(25,25){\vector(1,1){0}}
\qbezier[20](65,35)(75,25)(90,10)\put(75,25){\vector(1,-1){0}}
\put(15,50){\makebox(0,0){\small \(1\)}}
\put(17,10){\makebox(0,0){\small \(2\)}}
\put(85,22){\makebox(0,0){\small \(4\)}}
\put(50,50){\makebox(0,0){\(L\)}}
\put(85,55){\makebox(0,0){\small \(3\)}}
\put(70,50){\line(1,0){30}}\put(85,50){\vector(1,0){0}}
\qbezier(32,60)(10,80)(10,50)
\qbezier(10,50)(10,20)(32,40)
\put(10,48){\vector(0,-1){0}}
\end{picture}
\end{center}
\caption{Partial trace of the node in Fig. \ref{node}.\label{partialtrace}}
\end{figure}

\noindent The relative position of the two nodes in the diagram has no relevance. All that is relevant is the joining of the two lines. As a particular case, the outgoing and incoming lines could belong to the same node. The node with such a joining of two of its lines we shall call the {\em partial trace\/} of the original node. The partial trace of \(L\) of Fig. \ref{node} is given by Fig. \ref{partialtrace}.
 It's useful to note that sometimes composition and partial trace can be performed after changing a line to its opposite, and we shall continue to designate these proceedures by the same words. \footnotetext{This is analogous to, and not totally disconnected from, particle physics {\em crossing relations\/} in which in a particle reaction one can pass some particles to the other side provided they get changed to the corresponding antiparticles. Thus proton-electron scattering \({\rm p+e^-\to p+ e^-}\) is related to proton-antiproton conversion  into an electron positron pair \({\rm p+\bar p\to e^-+e^+}\). Other types of processes such as \({\rm p\to p+e^-+e^+}\), etc. can't take place in free space by conservation laws but can happen in strong background fields. In relativistic field theory the PCT symmetry\cite{stre-wigh} establishes a linear isomorphism between the physical Hilbert space and it's dual, pushing thus the distinction between the \physical\ space and the \virtual\ one under the rug.}

An explicit definition of the composite, which is an element of \linebreak \(\cL(\cH_1\otimes \cH_2^*\otimes \cH_5^*, \cH_1\otimes \cH_2^*\otimes \cH_4^*)\), is given using tensor universality as follows. An element such as \(L\in \cL( \cH_1\otimes \cH_2^*, \cH_1\otimes \cH_3\otimes \cH_4^*)\) is a sum of elements of the form \(A\otimes e\) where \(e\in \cH_3\) and \(A\in \cL( \cH_2^*\otimes \cH_1, \cH_1\otimes \cH_4^*)\), and an element such as \(M\in \cL(\cH_3\otimes \cH_5^*, \cH_2)\) is a sum of elements of the form \((f,\cdot)\otimes B\), where \(f\in \cH_3\) and \(B\in \cL(\cH_5^*,\cH_2)\). For these two elements we define the composite as \((f,e)A\otimes B\) which by tensor universality uniquely defines the composite in general. Note that by expressing \(e\) in \(A\otimes e\) as a function on \(\cH_3^*\) we get \(A\otimes (\cdot ,e)\) and naming \((f,\cdot)\) in \((f,\cdot)\otimes B\) we get \(\bar f\otimes B\). These are now objects in \(\cL( \cH_2^*\otimes \cH_1\otimes \cH_3, \cH_1\otimes \cH_4^*)\) and \(\cL( \cH_5^*,\cH_3^*\otimes \cH_2)\) respectively but whose composite is again \((f,e)A\otimes B\).  This corresponds to changing the line of \(\cH_3\) in Fig. \ref{nodecomposition} to its opposite  (opposite direction and \essence). Such a change on lines that leave from and terminate on a node do not change the overall object.

The partial trace is defined similarly. Any element of \(\cL( \cH_1\otimes \cH_2^*, \cH_1\otimes \cH_3\otimes \cH_4^*)\) is a sum of elements of the form \((e,\cdot)\otimes B\otimes f\) with \(e,\,f\in \cH_1\) and \(B\in \cL(\cH_2^*,\cH_3\otimes \cH_4^*)\). The partial trace of this element is \((e,f)B\) and tensor universality takes care of the rest.

It's now time to introduce an alternative formalism for the same objects, which we borrow from general relativity. Readers familiar with tensor and spinor formalisms in general relativity can skim this part, though not skip it altogether as some relevant points are made here and there. Among the  main objects here are  tensor fields on the space-time manifold \(M\). At each space-time point \(p\)  a tensor field is an  element of a finite tensor product of some copies of the tangent space \(T_pM\) and it's dual, the cotangent space \(T^*_pM\), at that point. For instance, the Riemann curvature tensor \(R(p)\) at \(p\) is an element of \(T_pM\otimes T^*_pM\otimes T^*_pM\otimes T^*_pM\). We shall now stop indicating the point \(p\) as it will be understood we're dealing with objects at a fixed point. If one has a coordinate system in a neighborhood of \(p\) one can introduce convenient bases for the tangent and cotangent space. The basis for the tangent space is denoted by \(\displaystyle \frac{\partial}{\partial x^i}\) where \(i=1,\dots, d\) with \(d\) being the dimension of space-time.\footnote{For most of us this is \(4\) but string theorists would disagree.} One then takes the dual basis for the cotangent space, which is denoted by \(dx^j\) and of course one has \(\displaystyle \left\langle dx^j,\frac{\partial}{\partial x^i}\right\rangle =\delta ^j_i\) where the Kronecker symbol \(\delta ^j_i\) is one if \(i=j\) and zero otherwise. Given these bases we can expand \(R\) in them and introduce its {\em components\/} \(R^i{}_{jk\ell}\) meaning that
\begin{equation}\label{repugnant}
  R = \sum_{i=1}^d\sum_{j=1}^d\sum_{k=1}^d\sum_{\ell=1}^d R^i{}_{jk\ell}\,\frac{\partial}{\partial x^i}\otimes\,dx^j\otimes\,dx^k\otimes\,dx^\ell.
\end{equation}
Note the systematic placement of indices, the upper indices on the components of \(R\) sum over the tangent space basis elements and the lower over the cotangent space basis elements. The bases themselves have indices placed opposite, with the proviso that an upper index in the denominator acts as a lower index.

All physicists of course find expression (\ref{repugnant}) repugnant\footnote{Many mathematicians still insist on writing (\ref{repugnant}).} and simply indicate the object \(R\) by its components \(R^i{}_{jk\ell}\) which we shall do from now on. Given another tensor \(S^p{}_q\) we can form the composite \(T^i{}_{jk\ell}\) given by
\begin{equation}\label{composite}
T^i{}_{jk\ell}=\sum_{n=1}^dS^i{}_nR^n{}_{jk\ell}.
\end{equation}
Physicist's find even (\ref{composite}) repugnant and adopt the {\em summation convention\/} by which an upper and a lower index that is repeated is summed over and so they simply write \(T^i{}_{jk\ell}=S^i{}_nR^n{}_{jk\ell}\) and we also do so from now on. Formula (\ref{composite}) is valid in any coordinate basis, expressing an intrinsic composition process which could have been equivalently defined by tensor universality, exactly the same way we did for the Hilbert space case above,  without recourse to a basis.

The use of tensor components in a basis along with the summation convention is a convenient formalism to do calculations, but lacks the basis independent aspect that mathematicians like so much and which in fact is quite important as one would like to know what results are basis independent and what are not. Fortunately there is a happy mean introduced by Wald\cite{wald:general.relativity} called the {\sl abstract index notation\/}. We denote the tensor \(R\) by \(R^a{}_{bcd}\) where by the latter symbol we do {\em not\/} mean the components of \(R\) in a basis but precisely the object \(R\). The indices indicate in which tensor product the object lies and also are useful to indicate which composites one forms with them. Composites are indicated by repeated indices, one upper one lower. We thus have \(T^a{}_{bcd}=S^a{}_eR^e{}_{bcd}\). Again, there is no sum here, the repeated indices simply indicate which composite is being formed. At times it {\em is\/} useful to consider the components of a tensor in a basis and Wald adopts the convention of using Greek indices for these, which we shall also do in the sequel. However in quantum mechanics one generally use Latin letters and when convenient, especially in referring to expression in the literature, we will do so also, and if needed for greater clarity also place the symbol in brackets; thus we would write \(R^\alpha {}_{\beta \gamma \delta }\)  or \([R]^m{}_{npq}\) for the components of \(R\).

There are certain additional structures in general relativity which are absent in the quantum mechanical tensor products. In general relativity there is a {\em metric tensor\/} \(g_{ab}\). As a matrix, \(g_{\alpha \beta }\), this is invertible, the inverse of which gives a tensor \(g^{ab}\). This can now be used to {\em raise\/} and {\em lower\/} indices by forming composites with these tensors. The tensor \(g_{ae}R^e{}_{bcd}\) is denoted by \(R_{abcd}\) and the tensor \(g^{bc}S^a{}_c\) is denoted by \(S^{ab}\). Notice that the name of the object does not change as the index is raised and lowered, something the mathematicians frown upon but which follows the time-honored physicists' tradition to use the same letter for a given physical quantity no matter how it is expressed. The metric tensor establishes a {\em linear\/} isomorphism between the tangent space and its dual: \(v^a\mapsto g_{ab}v^b\).

In quantum mechanics there is nothing corresponding to the metric tensor and there is no canonical linear isomorphism between a Hilbert space and its dual. There is however the Riesz duality and the isomorphism given by (\ref{caniso}).
We shall capture these notions in an abstract index notation for quantum mechanics. A close relative of this can be found in the {\em spinor\/} formalism of general relativity\cite{wald:general.relativity,penrose:AP10.171}. In this context, a spinor at a point in \(M\) is an element of a complex vector space acted upon by the regular representation of \(\hbox{SL}(2,\lC)\). It is thus a two-component object. One also considers elements in a space acted upon by the adjoint representation, which should be called {\em cospinors\/} but in the literature are not. {\em Spinorial tensors\/} are objects in tensor products of such spaces. Forgetting the original context, these objects are elements of tensor products of two-dimensional complex spaces, and so the algebra of spinorial tensors in general relativity is very similar to the algebra of multipartite qubit systems. One can make s few notational bridges between the two.

In the spinorial tensor algebra there is method of raising and lowering indices but this is very dependent on the relevant representations being two-dimensional and on space-time being four-dimensional which does not translate to the general quantum mechanical situation. We'll thus briefly consider only lower indices. In the index notation for a spinorial tensor an index that corresponds to the adjoint representation is dotted (or primed as in \cite{wald:general.relativity}). Thus \(\psi _{AB\dot C}\) is, using the abstract index notation, an element of the tensor product of two copies of the space of the regular representation and one of the adjoint. One can pass to the adjoint representation by going over to the dual space. 
We are then exactly in the situation described previously in our prelude to the quantum mechanical prop.

For an element such as \(L\) of Fig. \ref{node} we thus introduce the expression
\begin{equation}\label{morph1}
  L_{a\bar b}{}^{cd\bar e}
\end{equation}
where lower indices indicate incoming lines, upper outgoing, and barred\footnote{I've found barring more convenient than dotting or priming as the bar also indicates complex conjugation.} indices \virtual\ spaces. This is {\em abstract index notation\/} and does not indicate components of the object in any basis. This symbol does not indicate which Hilbert space each index refers to but keeping this information always present would make the notation cumbersome, so we'll adopt  two conventions for dealing with this: (I) If the object is first given by explicitly stating that it is an element of \(\cL(P,Q)\) then the indices are first the lower ones in the order of the spaces in \(P\), and then the upper ones in the order of the spaces in \(Q\).\footnote{This is opposite to the convention used for matrices but is my personal strike against the absurd practice of composing maps left to right diagrammatically and right to left notationally. If \(X\stackrel{f}{\to} Y\stackrel{g}{\to}Z\), then \(f(x)\) ought to be written \((x)f\) and \(g\circ f\) ought to be written \(f\circ g\), oughtn't they?} This is what was done in (\ref{morph1}). (II) Otherwise, the Hilbert space will be indicated in the blank spaces above or below the corresponding abstract index. This will be done when the object is first presented, and then this extra information will be removed. Thus (\ref{morph1}) would according to this convention be first presented as \(L_{a\bar b}^{12}{}^{cd\bar e}_{134}\), and subsequently the numbers will be absent.
Each index thus has two attributes, it's {\em position\/}, either {\em lower\/} or {\em upper\/}, which in the diagrammatic formalism correspond to either incoming or outgoing line, and its {\em \conjugacy\/}, either {\em unbarred\/} or {\em barred\/}, which in the diagrammatic formalism correspond to either solid or a dotted line.

It's useful to give a name to the objects represented by this abstract index notation. Neither ``tensor" nor ``spinor" is quite adequate as neither expresses the true propic\footnote{Even so, general relativists have been working with a prop structure for the better part of a century.} nature of them. We've  settled on ``morph" partially because they can be interpreted in many ways as morphisms. This will be clear soon.

We can already state some basic properties of morphs.
\begin{enumerate}
 \item Any lower index can be raised and any upper lowered provided we change its \conjugacy. Thus the two versions of \(L\) in Fig. \ref{nodeversions} are \(L^{\bar a b c d\bar e}\) and \(L_a{}^{bcd}{}_e\). Note that we continue to use the same name ``\(L\)" for these new objects in conformity with the practice in general relativity.

\item \label{moprhcomp} Any two morphs can be composed provided some upper indices of one refer to the same Hilbert spaces as some lower indices of the other, and the corresponding indices have the same \conjugacy.   Composition is indicated by repeating the indices involved. Thus the composite of \(L\) of Fig. \ref{node} with \(M_{a\bar b}^{35}{}^e_2\) of Fig. \ref{anothernode} is indicated by \(L_{a\bar b}{}^{cp\bar e}M_{p\bar f}{}^g\).

\item One can form a partial trace of a morph  provided it has an upper and a lower index of the same \conjugacy\ referring to the same Hilbert space. The partial trace is indicated by repeating the index. Thus one has \(L_{a\bar b}{}^{ad\bar e}\) for the morph of Fig. \ref{node} whose partial trace is depicted in Fig. \ref{partialtrace}.

\item \label{idmorph}For any Hilbert space, \physical\ or \virtual, there is an {\em identity moph\/} corresponding to the identity map and which we denote by \(\delta_a{}^b\), or \(\delta _{\bar a}{}^{\bar b}\) respectively. One of course has \(\delta_b{}^a M^{\cdots b \cdots}{}_{\cdots}=M^{\cdots a \cdots}{}_{\cdots}\) and \(\delta_a{}^c R^{\cdots}{}_{\cdots c \cdots}=R^{\cdots }{}_{\cdots a \cdots}\) for any \(M\) and \(R\) and analogously for \(\delta _{\bar a}{}^{\bar b}\).

\item Complex numbers are to be considered morphs. They have no abstract indices.
\end{enumerate}

We see from item (\ref{moprhcomp}) that composing, in the same way,  a given fixed morph with morphs of a fixed other space defines a morphism between two tensor product spaces. A given morph can thus defines infinitely many morphisms between pairs of tensor product spaces. This not only points out it's propic nature, as opposed to merely categorical, but also shows that the name ``morph" is apt. Also, ever since computer graphics became commonplace, ``to morph" means also to change shape. This is also apt as our morphs change their shape (index placements) when called upon to play different roles. We shall use {\em morphic\/} as the adjective corresponding to {\em morph\/}.\footnote{I also considered {\em morphetic} to rhyme with {\em prophetic} but thought that sounded too smug.}

As mentioned before, each morph is also an element of a Hilbert space with the inner product given by (\ref{inner}). In the Hilbert space prop therefore the distinction between object and morphism is largely dissolved, again in contrast to the merely categorical view. Each morph is thus also the Riesz representative of an element in the dual space. We denote the element it represents by barring the symbol and in the diagrammatic  representation also change each line \essence\ to its opposite while in the morph formalism change the \conjugacy\ of each index. Thus the Riesz conjugate of  \(L\) of Fig. \ref{node} is diagrammatically given by Fig. \ref{nodebar} and its morph notation is \( \bar L_{\bar a b}{}^{\bar c\bar d e}\).

\newsavebox{\nodebar}
\savebox{\nodebar}{
\put(50,50){\circle{40}}
\qbezier[20](0,40)(15,45)(30,50)\put(15,45){\vector(3,1){0}}
\put(10,10){\line(1,1){25}}\put(25,25){\vector(1,1){0}}
\put(65,35){\line(1,-1){25}}\put(75,25){\vector(1,-1){0}}
\qbezier[20](35,65)(22,78)(10,90)
\put(25,75){\vector(-1,1){0}}
\put(15,40){\makebox(0,0){\small \(1\)}}
\put(17,10){\makebox(0,0){\small \(2\)}}
\put(33,75){\makebox(0,0){\small \(1\)}}
\put(85,22){\makebox(0,0){\small \(4\)}}
\put(50,50){\makebox(0,0){\(\bar L\)}}
}

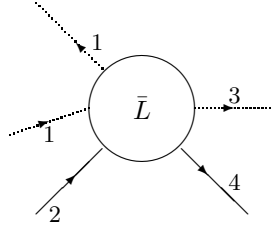
\begin{figure}[h!]
\begin{center}
\begin{picture}(100,100)(0,0)
\put(0,0){\usebox{\nodebar}}
\put(85,55){\makebox(0,0){\small \(3\)}}
\qbezier[20](70,50)(85,50)(100,50)
\put(85,50){\vector(1,0){0}}
\end{picture}
\end{center}
\caption{Riesz conjugate of \(L\) in Fig.\ref{node} \label{nodebar}}
\end{figure}

If \(Q\) is another morph from the same space as \(L\) then its inner product with \(L\) is
\begin{equation}\label{innermorph}
  (Q, L)=\bar Q^{a\bar b}{}_{cd\bar e}L_{a\bar b}{}^{cd\bar e}.
\end{equation}

Note that the morph \(\bar Q\) on the right-hand side is obtained from the Riesz conjugate \( \bar Q_{\bar a b}{}^{\bar c\bar d e}\) by changing each index to its opposite (changing its position and \conjugacy). In this role the morph repreresents the {\em adjoint\/} of \(Q\). This is the morphic version of defining the adjoint as the transpose of the complex conjugate. Any morph that has no free indices, that is all indices belong to repeated pairs, represents a complex number. To calculate the complex conjugate of this number one simply bars everything (morph symbols and indices). Thus the complex conjugate of the inner product (\ref{innermorph}) is \(Q^{\bar ab}{}_{\bar c\bar de}\bar L_{\bar ab}{}^{\bar c\bar d e}\). Bar of a bar is of course nothing.

In accordance with item \ref{idmorph} one has \(\bar \delta _{\bar a}{}^{\bar b}= \delta _{\bar a}{}^{\bar b}\) the right-hand side being previously defined in the item referred to. Bars over the symbol \(\delta \) can therefore be dropped.

\begin{figure}[h!]
\begin{center}
\includegraphics[scale =0.3]{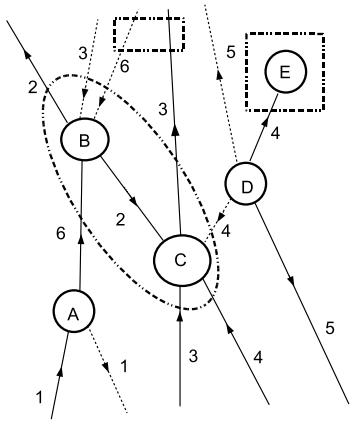}
\end{center}
\caption{Contour of a new node \(F\).\label{boxes}}
\end{figure}

\begin{figure}[h!]
\begin{center}
\includegraphics[scale =0.3]{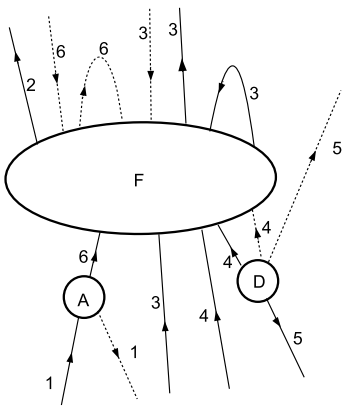}
\end{center}
\caption{Reduced diagram. \label{redboxes}}
\end{figure}

We're now ready to explain how morphs form a mathematical prop. Given any diagram we can enclose any part of it by a closed contour, connected or not, provided one does not cut through any node. Thus in Fig. \ref{boxes} the oval and the two rectangles are such a contour. The contour can then be thought of as a new node and the diagram reduced by omitting
 the  part of the diagram within and expressing the node in the usual way as a closed curve with one component. For our example, labeling the new node by \(F\), this results in Fig. \ref{redboxes}.

In terms of morph composition the diagram of Fig. \ref{boxes} is given by:
 \[A_{a \bar b}^{11}{}_6^cB_{c \bar d\bar e}^{636}{}_{22}^{fg}C_{ghi\bar \jmath}^{2344}{}_3^kD_{4554}^{\bar \jmath\bar \ell mn} E_n^4.\]

Forming a new node is juxtaposing morph symbols together and considering such a juxtaposition  as defining a new morph. A use of identity morph symbols is sometimes necessary to account for lines that enter and leave the contour without encountering a node. Thus Fig. \ref{redboxes} in morphic terms is

\[A_{a \bar b}{}^cD^{\bar \jmath\bar \ell mn}\{B_{c \bar d\bar p}{}^{fg}C_{gqi\bar \jmath}{}^k E_n \delta_{\bar e}{}^{\bar p}\delta _h{}^q \},\]
where the expression in braces is
\[F_{c\bar d\bar pgqi\bar\jmath n\bar e h}{}^{fgk\bar p q}.\]
There are two pairs of contracted indices in this expression which in Fig. \ref{redboxes} correspond to lines that leave and return to node \(F\), these are partial traces.

One can continue with this process and introduce contours in the reduced diagram, reduce this, and repeat any number of times. In terms of the original unreduced diagram this corresponds to adding new contours with the proviso than any that are already present must be wholly within or wholly without the new ones. The defining property of a prop is now the following: any contour which is within another can be eliminated without changing the object. This is a form of associativity of composition appropriate to many-valued many-variable maps. We shall not give the precise formal expression for this associativity as the intuitive idea is quite clear. In  a category  a node (which is not an object but a morphism) must have exactly one incoming and one outgoing line and no compositional loops are allowed so all diagrams are just vines (in an operad they are trees, props are more tropical), the corresponding associativity condition reduces to the one in the Fig \ref{catassoc}:
\begin{figure}[h!]
\begin{center}
\includegraphics[scale =0.5]{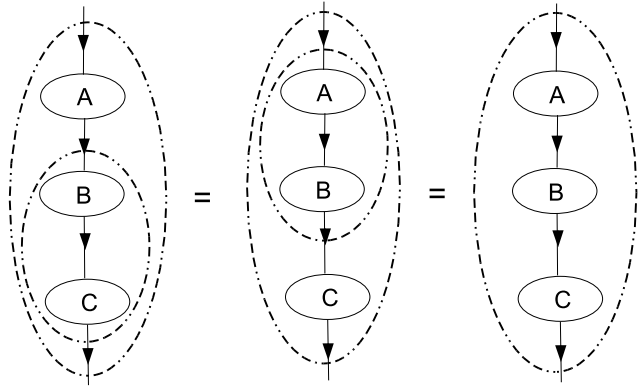}
\end{center}
\caption{Categorical associativity. \label{catassoc}}
\end{figure}

Summing up, one now has four notational ways to deal with quantum information systems. The traditional physicists' way using Dirac's bra-ket notation, the mathematicians' way with their traditional symbology, the diagrammatical way, borrowed from many sources both physical and mathematical, and now, the morphic way borrowed from general relativity.\footnote{After having thought of this notation, I discovered that some authors in Eastern Europe have already been using it. Unfortunately I've not been able to discover who introduced it for the first time. John Baez\cite{baez:universal} also makes a connection between  morphs, Feynman diagrams and tensors, though there's no distinction between \physical\ and \virtual\ Hilbert spaces. Czachor \cite{czac:CQG25.205003} analyzes teleportation with essentially an identical formalism.} The diagrammatical way introduced here differs from many others in the literature mainly by distinguishing \physical\ and \virtual\ Hilbert spaces as suggested by (\ref{caniso}). Hybrid schemes also abound as each way has its advantages and shortcomings and often one combines the better aspects of several. The two presented in this section have the advantage of allowing, in a natural way, base independent constructs, calculations and proofs. The diagrammatical method is useful for expressing various mental props while the morphic provides concise calculations, once one gets used to manipulating many indices. These facts will be illustrated the next sections.

\section{Base independence and dependence}\label{indep}

Base independence in quantum information is like coordinate independence in general relativity. The physics does not depend on bases or coordinates but physicists' activities do as they need  to record their observations and communicate them to others.
Laboratory props generally determine useful bases for describing results. One needs a convenient way to go from a base-independent formalism to a base related one, and back. Once again, we'll borrow from general relativity. In morphic terms an orthonormal base in a Hilbert space is a set of morphs \(e_\alpha{}^a\) where \(\alpha =1,\dots,d\) and \(d\) is the dimension of the Hilbert space.  The orthonormality condition is
$e_\alpha{}^a\bar e_\beta{}_a=\delta _{\alpha \beta }.$
 In four dimensional relativity such an object (at a point) is called a {\em vier-bein\/} from the German words {\em vier\/}\footnote{Pronounced exactly like the English word {\em fear}.} meaning four and {\em bein\/}\footnote{Pronounced {\em buy-n}.} meaning leg. This is an orthonormal basis for the tangent space and is thus seen as a four-legged beast. In \(n\)-dimensional relativity\footnote{String theorists loved \(n=26\) but then went for someone younger, \(n=10\), which seems to be ageing toward \(n=11\).} such an object is called an {\em \(n\)-bein\/}. Since quantum informaticists\footnote{``There ain't no such word!" you say. I say ``Google it!" Actually, a good catchy term for a quantum information specialist is still lacking. I think {\em qdude} for a man is just groovy (if you don't understand this, you're much younger than I) but I'm not sure what to use for a woman. How about {\em qfemme?} For the plural any of {\em qdudes\/}, {\em qguys\/} or {\em qcats\/} will do.} often use \(d\) for Hilbert space dimension (behold the ``qdit") we will call such a collection of morphs a {\em \db}. Unless certain precautions are taken, the introduction of components in a basis can lead to much confusion and obfuscation of essential aspects. The dual basis of  \(e_\alpha{}^a\)  is of course  \(\bar e_\alpha{}^{\bar a}\)  whose other morphic form was used in the orthonormality condition. Morph components are of course obtained by composing with \db{s}. Thus the components of \(A_a{}^{\bar b}\) would be given by:
$ A_{\alpha \beta }= A_a{}^{\bar b}e_\alpha{}^ae_\beta{}_{\bar b}$
using the two morphic forms of the \db. The result is however ambiguous as \(A_a{}^be_\alpha{}^a\bar e_\beta{}_b\) gives the same expression but the original morphs are different. One cannot determine if the composition was with a \db\ or its Riesz dual. To circumvent this ambiguities and to add greater flexibility to the notation we find it convenient to bar or not a component index and also to allow raising and lowering it with the proviso that if one changes its position one must also changes its duality. All ambiguities are now resolved provided we denote the Riesz dual of \(e_\alpha{}^a\) by  \(\bar e_{\bar\alpha}{}^{\bar a}\) and decree that all other morphic forms of a \db\ are now obtained from these two by the rules of raising, lowering, and barring. One must of course rewrite the orthonormality condition as:
\begin{equation}\label{ortobein}
e_\alpha{}^a\bar e_{\bar\beta}{}_a=\delta _{\alpha \bar\beta }.
\end{equation}

Barred indices of course assume the same range of values as unbarred ones. The bar is only a device that keeps track of the nature of the original morph.
Given the components of a morph, for example \(S^\alpha {}_{\beta \bar\gamma }\), the reconstruction of the original morph in abstract index notation is not unique but the various forms differ only by raising and lowering the indices with the rules for doing so maintained.

We shall call the attribute of an index that determines whether it is abstract or component its {\em species\/}.

Since introducing morph components is nothing but composing with members of a \db, one can mix abstract and component indices, thus from \(S\hlu{a}{b\bar c}\) one can form for instance \(S\hlu{a}{\beta \bar c}=S\hlu{a}{b\bar c}e\hlu{\beta }{b}\). What \(S\hlu{a}{\beta \bar c}\) means is that for any value of the index \(\beta \) one has a morph of the form \(M\hlu{a}{\bar c}\). In the same vein, \(S^\alpha {}_{\beta \bar\gamma }\) means that for all values of the component indices one has a morph with no abstract indices at all, that is, a complex number. We shall call objects with such mixtures of indices {\em hybrid morphs\/}.

It's easy to see that, by the very definition of an orthonormal basis,
\begin{equation}\label{sumbein}
  \sum_{\alpha =1}^d e\hlu{\alpha }{a}\bar e^\alpha {}_b=\delta\hlu ab.
\end{equation}

We shall of course adopt a {\em summation convention\/} for a  repeated component index in opposite positions, denoting summation over all its values. Thus (\ref{sumbein}) now becomes \(e\hlu{\alpha }{a}\bar e^\alpha {}_b=\delta\hlu ab\). The Kronecker symbols \(\delta _{\alpha \bar \beta }\), \(\delta \hlu\alpha \beta \) etc.,  thus give the components of the identity morph. Recovery of the original morph from its components is now seen to be given by summing over the components and the indices of a product of \db{s}. Thus:
\begin{equation}\label{backtomorph}
S\hlu a{b\bar c}=S^\alpha {}_{\beta \bar\gamma }e^\alpha{}_a\bar e^\beta {}_b e^{\bar \gamma }{}_{\bar c}.
\end{equation}

What (\ref{ortobein}) and (\ref{sumbein}) mean is that a repeated abstract index can be changed to a repeated component index, and vice-versa, without changing the object. Thus one has the hybrid morph relation \(A\hlu{a\beta }{\gamma }B\hlu{\gamma \delta }{}= A\hlu{a\beta }{c }B\hlu{c \delta }{}\). Changing the species of any other index will give an equivalent relation but the nature of the composite object changes also.

We are now ready to illustrate the usefulness of the above formalism by examining a few familiar circumstances in quantum information.
\vskip 24pt

\subsection{Partial transpose.}

Consider a bipartite density matrix \(\rho \) in \(\cH_1\otimes \cH_2\). Admitting orthonormal bases for the spaces one generally represents \(\rho \) by a {\em matrix\/} with composite indices thus: \(\rho _{ij;nm}\). Its {\em partial transpose\/} is then given by
\begin{equation}
  \rho^t _{ij;nm}= \rho _{nj;im}.
\end{equation}
This is obviously a basis-dependent construct, however it's shown to be an important one in quantum information theory. Entangled density matrices for which the partial transpose is a positive matrix are known as {\em bound entangled\/} an important and rather enigmatic class. Surely there must be some base-independent way to define them, seeing that {\em positivity\/} of the partial transpose is a base-independent property.

In morphic terms, the density matrix, in its role as an element of \(\cL(\cH_1\otimes \cH_2,\cH_1\otimes \cH_2)\), is given by \(\rho \hlu{ab}{cd}\). It plays another role through (\ref{caniso}) as an element of \(\cL(\cH_1^*\otimes \cH_2,\cH_1^*\otimes \cH_2)\) and which is given by \(\rho ^{\bar a}{}_{b \bar c}{}^d\), raising one index and lowering another. {\em This is the base-independent partial transpose.} Diagrammatically this is depicted in the following figure.

\begin{center}
\begin{picture}(120,80)
\put(00,30){\framebox(40,20)}
\put(80,30){\framebox(40,20)}
\put(10,0){\line(0,1){30}}
\put(30,0){\line(0,1){30}}
\put(110,0){\line(0,1){30}}
\put(10,80){\line(0,-1){30}}
\put(30,80){\line(0,-1){30}}
\put(110,80){\line(0,-1){30}}
\put(10,18){\vector(0,1){0}}
\put(30,18){\vector(0,1){0}}
\put(110,18){\vector(0,1){0}}
\put(10,68){\vector(0,1){0}}
\put(30,68){\vector(0,1){0}}
\put(110,68){\vector(0,1){0}}
\put(90,13){\vector(0,-1){0}}
\put(90,63){\vector(0,-1){0}}
\qbezier[15](90,0)(90,15)(90,30)
\qbezier[15](90,50)(90,65)(90,80)
\put(20,40){\makebox(0,0){\(\rho \)}}
\put(100,40){\makebox(0,0){\(\rho \)}}
\end{picture}
\end{center}

We now show that the positivity of this object is equivalent to the positivity of the conventional partial transpose, thereby also proving the base-independence of the property.
We must establish the correspondence between the morphs and the conventional matrix elements of \(\rho \) and \(\rho ^t\). Since the conventional matrix elements are indexed by Latin letters, we use the bracket convention and write \([\rho ]\hlu{nm}{ij}\) for the the components of the density morph. The bracket is the morph's way of saying ``All  those indices are Greek to me!" By our convention, lower indices are inputs and upper are output, and matrix convention is that input indices are on the right and output on the left, so we've established:
\[\rho _{ij;nm}=[\rho ]\hlu{nm}{ij}.\]
The corresponding components of the base-independent partial transpose are \([\rho] ^{\bar n}{}_{m \bar \imath}{}^j\). Taking into account which index belongs to which Hilbert space, and which is incoming and which outgoing, and after dropping the bars which are not used in matrix notation, the corresponding object is exactly the conventional partial transpose, thus:
\begin{equation}\label{badstep}
 \rho^t _{ij;nm}=[\rho] ^{\bar \imath}{}_{m \bar n}{}^j.
\end{equation}

The positivity of \(\rho ^t\) means that for all martices \(C_{rs}\) of complex numbers one has \(\sum_{ijnm}C_{ij}^*\rho^t _{ij;nm}C_{nm}\ge 0\). Here, for the moment, we've suspended the summation convention and index placement as upper or lower, and used the asterisk to denotes complex conjugation. In other words:
\[\forall\,C_{rs},\,\,\sum_{ijnm}C_{ij}^*[\rho] ^{\bar \imath}{}_{m \bar n}{}^jC_{nm}\ge 0.\]

Now any complex matrix \(C_{rs}\) can be obtained as the components of a morph in the role of an element of \(\cH_1\otimes \cH_1^*\) thusly: \(C_{rs}=[\Phi ]^{s\bar r}\). This is clear as the two spaces have the same dimension. One then has \(C_{rs}^*=[\bar\Phi ]_{\bar sr}\). Positivity of \(\rho ^t\) now is equivalent to
\[\forall\,\Phi ^{a\bar b},\,\,\sum_{ijnm}[\bar\Phi ]_{\bar \imath j}[\rho] ^{\bar \imath}{}_{m \bar n}{}^j[\Phi ]^{\bar m n}\ge 0.\]
We can now readopt the summation conventions and change all indices (secretly Greek, by the bracket  convention) to abstract ones to arrive at
\[\forall\,\Phi ^{a \bar b},\,\,\bar\Phi_{\bar c d}\,\rho ^{\bar c}{}_{e \bar f}{}^d\,\Phi ^{e\bar f}\ge 0.\]

This is precisely the positivity of \(\rho ^{\bar a}{}_{b \bar c}{}^d\) in its role as an element of \(\cL(\cH_1^*\otimes \cH_2)\) and is a base independent statement.

In going from the the right-hand side of (\ref{badstep}) to the left-hand side we've lost track of the morphic nature of the density matrix by neglecting the barred indices and by forcing it into the straight-jacket of a morphism in an impoverished category.\footnote{Too much {\em depropification\/}(a direct generalization of the notion of decategorification\cite{baez:decat}) has occurred.} After this, recovering any base-independent conclusions becomes exceedingly arduous.

Another place where the partial transpose appears is in the {\em Jamio\l{kowski} criterion\/} concerning completely positive maps. We shall deal with several concepts of positivity for operators. Recall that the usual one for an operator \(A\) on a Hilbert space \(\cH\) is that \((\Phi ,A\Phi )\ge\) for all elements \(\Phi \in \cH\)\footnote{Strictly speaking, one should call such an operator {\em non-negative\/} but {\em positive\/} seems to be the prevailing term.}, and we denote this by \(A\ge 0\). This notion of course is what was involved in the discussion above about the partial transpose of a density matrix. Other notions of positivity will have a qualifying adjective. The set of positive operators is a {\em positive cone\/}. This means that if \(A,\,B\) are positive operators and \(r,\,s\) are non-negative real numbers then \(rA+sB\) is positive, and  that if \(A\) and \(-A\) are positive, then \(A=0\).

If now \(\Lambda: \cL(\cH)\to \cL(\cH)\) then we say that it is {\em cone-positive\/} if it maps the positive real cone of positive operators into itself. That is \(A\ge 0\Rightarrow \Lambda (A)\ge 0\). Being positive and cone-positive are two different things. For instance, if \(B\ge 0\) then \(\Lambda (A)=BA\) defines a positive map since \(\Tr(A^*BA)\ge 0\) for all \(A\), but it is cone-positive only if \(B\) is a multiple of the identity. Likewise for any  \(M\), \(\Lambda  (A)=M^*AM\) is cone-positive but is  positive if and only if \(M=zB\) for some complex number \(z\) and \(B\ge 0\).

The cone-positive maps form a positive cone in \(\cL(\cL(\cH))\). A map \(\Lambda \) is called {\em completely positive\/} if for any Hilbert space \(\cK\) the map \(I_{\cL(\cK)}\otimes \Lambda \), considered as a map from \(\cL(\cK)\otimes \cL(H)\simeq \cL(\cK\otimes \cH)\)  to itself is cone-positive. Completely positive maps are important because they describe {\em quantum channels\/} in the sense that any physically realizable transformation of a density matrix is of the form \(\rho \mapsto \Lambda (\rho )\) where \(\Lambda \) is completely positive.\footnote{There are many arguments for this, the simplest one is that the result of a physical process on a system should not be affected by the existence of any other uncorrelated and uncoupled system. There are variations on this theme.} Given a base \(e_i\), \(i=1,\dots,d\)  in \(\cH\)  one defines \(E_{ij}=e_j(e_i,\cdot)\) and forms the object
\begin{equation}\label{jef}
\cJ_{e}(\Lambda )=\sum_{ij}\Lambda (E_{ij})\otimes E_{ij}
\end{equation}
understood as an element of \(\cL(\cH\otimes \cH)\). The criterion is:  \(\Lambda \) is completely positive if and only if \(\cJ_{e}(\Lambda )\) is positive. Obviously \(\cJ_{e}\), by its construction, is a base-dependent object, whereas complete positivity is a base-independent property. Surely there must be a similar criterion that is fully base-independent. In morphic terms \(A\in \cL(\cH)\) is given by \(A_a{}^b\) and so \(\Lambda (A)\) is given by \(\Lambda _a{}^{bc}{}_dA_c{}^d\) where \(\Lambda _a{}^{bc}{}_d\) is the morph playing the role of an element of \(\cL(\cL(\cH))\). It plays another role
\begin{equation}\label{goodj}
\Lambda \hlu ab{}\hlu {\bar c}{\bar d}
\end{equation}
as an element of \(\cL(\cH\otimes \cH^*)\) and which we'll refer to as \(\cJ(\Lambda )\). {\em This is the base-independent version of (\ref{jef}).} It is also another example of a partial transpose. It's instructive to present \(\cJ(\Lambda )\) in terms similar to (\ref{jef}), thus:
\begin{equation}\label{jok}
\cJ(\Lambda )=\sum_{ij}\Lambda (E_{ij})\otimes (\cdot,e_i)\bar e_j.
\end{equation}
This is the base-independent object written out in conventional notation. A choice of a base is necessary to {\em express\/} the object this way, but the result (the resulting sum) is independent of the choice. In morph notation no choice of bases is necessary. The Jamio\l{kowski} criterion is now: \(\Lambda \) is completely positive if and only if \(\cJ(\Lambda )\) is positive. This is easy to prove. Suppose \(\Lambda \) is completely positive, then it is known that it has a {\em Kraus representation\/} \(\Lambda (A)=\sum_{i=1}^kM_k^*AM_k\) for some maps \(M_k\in\cL(\cH)\), known as {\em Kraus maps\/}. Since completely positive maps form a positive cone, it's enough to prove necessity for the case one Kraus map. In this case we have \(\Lambda _a{}^{bc}{}_d=\bar M^b{}_dM\hlu ac\) and \(\cJ(\Lambda )\) is \(\bar M^{b\bar d}M_{a \bar c}\). If \(\Phi ^{a\bar c}\) represents an element of \(\cH\otimes \cH^*\), then \((\Phi ,\cJ(\Lambda )\Phi )=\bar\Phi _{b \bar d}M^{b\bar d}M_{a \bar c}\Phi ^{a \bar c}=\left|M_{a \bar c}\Phi ^{a \bar c}\right|^2\ge 0\) proving the positivity of \(\cJ(\Lambda )\). To prove sufficiency, suppose \(\cJ(\Lambda )\) is positive, then one has for \(\Phi ^{a\bar c}=\alpha ^a\beta ^{\bar c}\) that \((\Phi ,\cJ(\Lambda )\Phi )=\bar \alpha _b\bar \beta _{\bar d}\Lambda \hlu ab{}\hlu {\bar c}{\bar d}\alpha ^a\beta ^{\bar c}\ge 0\). The morph expression for this inner product can now be changed (by rasing and lowering indices) to \(\bar \alpha _b\{\Lambda_a{}^{bc}{}_d\bar \beta ^{d}\bar \beta _c\}\alpha ^a\) which can be seen to be precisely \((\alpha ,\Lambda ((\bar \beta ,\cdot)\bar \beta )\alpha )\), which is positive by hypothesis. Now \(\alpha \) is an arbitrary element of \(\cH\) and \((\bar \beta ,\cdot)\bar \beta \) is an arbitrary rank-one self-adjoint operator in \(\cH\) (remember that \(\beta \in \cH^*\)). Since any positive operator is a sum with positive coefficients of such rank-one operators we've proven that \(\Lambda \) is cone-positive. Now \(\cJ(\Lambda \otimes M)\simeq \cJ(\Lambda )\otimes \cJ(M)\). From the morph perspective this is obvious, since tensoring is juxtaposition and \(\cJ\) raises and lowers indices, and it's clear that juxtaposing and then moving indices is the same as moving indices and then juxtaposing. Now as a morph, \(I_{\cL(\cK)}\) is \(\delta _a{}^c\delta ^b{}_d\) (indices to be identified with those of \(\Lambda _a{}^{bc}{}_d\)), and so \(\cJ(I_{\cL(\cK)})\) is \(\delta _{a \bar c}\delta ^{b\bar d}\). This is a rank-one positive operator on \(\cH\otimes\cH^* \), precisely \(d\) times the orthogonal projection onto the subspace generated by  the canonical image of \(I_{\cH}\) in \(\cH\otimes\cH^* \) by (\ref{caniso}). The tensor product of two positive operators is positive, so \( \cJ(I_{\cL(\cK)})\otimes\cJ(\Lambda )\) is positive, and by what was proven above \(I_{\cL(\cK)}\otimes \Lambda \) is cone-positive and so we conclude that \(\Lambda \) is completely positive.

\subsection{The no-signaling theorem}

Alice and Bob each share one part of a bipartite state \(\Phi \in \cH_1\otimes\cH_2\). Alice couples her part to an ancillary state (the famous ancilla\footnote{{\em Almost\/} rhymes with {\em Godzilla}, too bad.}) with a unitary transformation. The no-signaling theorem, in one of it's many manifestations, says that Bob cannot know of Alice's actions by measurements performed on his part of \(\Phi \) for otherwise one could set up a superluminal communication device. Concretely this means that the density matrix obtained by a partial trace on the ancilla state and Alice's part is the same as the one obtained by just a partial trace on Alice's part before coupling to the ancilla. In Dirac notation the density matrix corresponding to a state \(\ket \Phi \) is \(\ket\Phi \bra\Phi \). This can be read in two ways, as an element \(\Phi \otimes \bar \Phi \in \cH\otimes \cH^*\)  or as a map \((\Phi, \cdot)\Phi \) from \(\cH\) to itself. Dirac notation is wonderfully ambivalent about this and one can choose to read it in the most convenient way at the moment. Diagrammatically the two versions are as in the following figure:
\begin{center}
\begin{picture}(130,50)
\put(0,0){\framebox(20,20)}
\put(30,0){\framebox(20,20)}
\put(80,0){\framebox(20,20)}
\put(110,0){\framebox(20,20)}
\put(10,10){\makebox(0,0){\(\Phi \)}}
\put(90,10){\makebox(0,0){\(\Phi \)}}
\put(40,10){\makebox(0,0){\(\bar \Phi \)}}
\put(120,10){\makebox(0,0){\(\bar \Phi \)}}
\put(10,20){\line(0,1){30}}
\put(90,20){\line(0,1){30}}
\put(120,20){\line(0,1){30}}
\qbezier[15](40,20)(40,35)(40,50)
\put(10,37){\vector(0,1){0}}
\put(40,37){\vector(0,1){0}}
\put(90,37){\vector(0,1){0}}
\put(120,31){\vector(0,-1){0}}
\end{picture}
\end{center}
In morphic terms these two versions are \(\Phi^a\bar\Phi ^{\bar b}\) and \(\Phi^a\bar\Phi_b\), less ambiguous than Dirac notation. However, when dealing with morphs one should not confuse  the actor and the character. The actor is the morph abstracting the position of the indices, and the character is the morph playing a given role with specific positions of the indices.

Returning to the no-signal theorem, the bipartite state is \(\Phi ^{ab}\), the ancilla is \(\Psi ^c\in \cK\), coupling to the ancilla one gets \(U_{ca}{}^{ef}\Psi ^c\Phi ^{ab}\). The corresponding density matrix, in one of its roles, is \(U_{ca}{}^{de}\Psi ^c\Phi ^{ab}\bar U_{\bar h\bar f}{}^{\bar \imath\bar\jmath}\bar\Psi ^{\bar h}\bar\Phi ^{\bar f\bar g}\). The partial trace is \(U_{ca}{}^{de}\Psi ^c\Phi ^{ab}\bar U_{\bar h\bar fde}\bar\Psi ^{\bar h}\bar\Phi ^{\bar f\bar g}\) lowering two indices on \(\bar U\) and repeating them with the corresponding ones on \(U\). Now \(U_{ca}{}^{de}\bar U_{\bar h\bar fde}=\delta _{c\bar h}\delta _{a\bar f}\) since \(U\) is unitary, and the result is \(\{\Psi ^c\bar \Psi _c\}\{\Phi ^{ab}\bar \Phi\hlu a{\bar g}\}\). The first factor is \(\|\Psi \|^2=1\), assuming the ancilla is normalized, and the second factor is the partial trace of the original density matrix. In contrast to other demonstrations,\footnote{For an early one for shared qbits and using Dirac bra-ket notation see \cite{benn-wies:PRL69.2881}.} no choice of basis was necessary.
The above demonstration includes the case of \(\Psi \) simply going along for the ride, that is \(U=I\otimes V\) where \(V\) is a unitary  in \(\cH_1\).

Note that if instead of  a unitary, one could use a general morph \(A\hlu{ab}{cd}\) the result would be false and Alice would be able to instantly communicate with Bob. Of course, Alice {\em can't\/} use a general operator and only certain morphs correspond to real laboratory props. One could in fact deduce the extent of Alice's possibilities by determining which operations do not allow for signals. Conventional wisdom is that all that one can do to a  density matrix is to subject it to a completely positive transformation, however neither the diagrammatic nor the morphic formalism is yet capable of handling this as such transformations are in general probabilistic when viewed in relation to individual states prepared in the laboratory.

It is instructive to perform the above calculation in diagrammatic form, the partial trace of the coupled ancilla and bipartite state density matrix is given by the followin figure:
\newsavebox{\nosig}
\savebox{\nosig}{
\put(0,0){\framebox(40,20)}
\put(110,0){\framebox(40,20)}
\put(20,90){\framebox(40,20)}
\put(90,90){\framebox(40,20)}
\put(40,50){\framebox(20,20)}
\put(90,50){\framebox(20,20)}
\put(20,130){\framebox(110,20)}
\put(20,10){\makebox(0,0){\(\Phi \)}}
\put(130,10){\makebox(0,0){\(\bar \Phi \)}}
\put(50,60){\makebox(0,0){\(\Psi \)}}
\put(100,60){\makebox(0,0){\(\bar \Psi \)}}
\put(40,100){\makebox(0,0){\(U\)}}
\put(110,100){\makebox(0,0){\(\bar U\)}}
\put(75,140){\makebox(0,0){\(I\)}}
\put(10,20){\line(0,1){150}}
\put(30,20){\line(0,1){70}}
\put(50,70){\line(0,1){20}}
\put(30,110){\line(0,1){20}}
\put(50,110){\line(0,1){20}}
\qbezier[110](140,20)(140,75)(140,170)
\put(10,62){\vector(0,1){0}}
\put(30,62){\vector(0,1){0}}
\put(30,122){\vector(0,1){0}}
\put(50,122){\vector(0,1){0}}
\put(50,82){\vector(0,1){0}}
\put(140,62){\vector(0,1){0}}
\put(5,35){\makebox(0,0){\(2\)}}
\put(35,35){\makebox(0,0){\(1\)}}
\put(115,35){\makebox(0,0){\(1\)}}
\put(145,35){\makebox(0,0){\(2\)}}
}
\begin{center}
\begin{picture}(150,170)
\put(0,0){\usebox{\nosig}}
\qbezier[15](100,70)(100,80)(100,90)
\qbezier[15](100,110)(100,120)(100,130)
\qbezier[15](120,110)(120,120)(120,130)
\qbezier[37](120,20)(120,55)(120,90)
\put(100,82){\vector(0,1){0}}
\put(120,62){\vector(0,1){0}}
\put(120,122){\vector(0,1){0}}
\put(100,122){\vector(0,1){0}}
\end{picture}
\end{center}

The \(I\) node is the identity morph in \(\cL(\cH_1\otimes \cK)\) in the role of the Riesz conjugate of it's role as an element of \(\cH_1\otimes \cK\otimes\cH_1^*\otimes \cK^*\).

We now change some of the arrows to their opposites, this is now given by the following figure:
\begin{center}
\begin{picture}(150,170)
\put(0,0){\usebox{\nosig}}
\put(100,70){\line(0,1){20}}
\put(100,110){\line(0,1){20}}
\put(120,110){\line(0,1){20}}
\put(120,20){\line(0,1){70}}
\put(100,78){\vector(0,-1){0}}
\put(120,58){\vector(0,-1){0}}
\put(120,118){\vector(0,-1){0}}
\put(100,118){\vector(0,-1){0}}
\end{picture}
\end{center}

The node \(\bar U\) with its lines changed to the opposite is the node for the adjoint \(U^*\) and so the nodes \(U\), \(I\), \(\bar U\), traversed in this order represent \(UIU^*=I\) and so can be replaced by the \(I\) node, but \(I_{\cH\otimes\cK}=I_{\cH}\otimes I_{\cK}\)  and so the sequence of nodes \(U,\,I,\,\bar U\) is just equivalent to having the incoming lines simply follow to the outgoing ones without intercepting anything. Thus the above diagram is equaivalent to this one:
\begin{center}
\begin{picture}(150,100)
\put(0,0){\framebox(40,20)}
\put(110,0){\framebox(40,20)}
\put(45,50){\framebox(20,20)}
\put(85,50){\framebox(20,20)}
\put(20,10){\makebox(0,0){\(\Phi \)}}
\put(130,10){\makebox(0,0){\(\bar \Phi \)}}
\put(55,60){\makebox(0,0){\(\Psi \)}}
\put(95,60){\makebox(0,0){\(\bar \Psi \)}}
\put(10,20){\line(0,1){80}}
\qbezier[110](140,20)(140,60)(140,100)
\put(10,62){\vector(0,1){0}}
\put(140,62){\vector(0,1){0}}
\put(5,35){\makebox(0,0){\(2\)}}
\put(35,35){\makebox(0,0){\(1\)}}
\put(115,35){\makebox(0,0){\(1\)}}
\put(145,35){\makebox(0,0){\(2\)}}
\put(65,60){\line(1,0){20}}
\put(77,60){\vector(1,0){0}}
\qbezier(30,20)(30,100)(75,100)
\qbezier(120,20)(120,100)(75,100)
\put(77,100){\vector(1,0){0}}
\end{picture}
\end{center}

The disconnected part in the middle is just the number \(\|\Psi \|^2=1\) and the rest is the partial trace of the density matrix before coupling to the ancilla. Note that in this partial trace, a line had to be changed to its opposite to be able to be joined to another one. The parallel to the morphic calculation is evident, but requires the work of drawing the diagrams. There's no advantage to this in this example, but in other cases below we'll see the real value of diagrammatic analysis.

\subsection{Coecke's theorem}

\begin{figure}[h!]
\begin{center}
\begin{picture}(65,96)
\put(5,0){\line(0,1){58}}\put(5,86){\line(0,1){10}}
\put(30,0){\line(0,1){15}}\put(30,43){\line(0,1){15}}
\put(30,86){\line(0,1){10}}
\put(55,0){\line(0,1){15}}\put(55,43){\line(0,1){53}}
\put(5,10){\vector(0,1){0}}\put(30,10){\vector(0,1){0}}
\put(5,95){\vector(0,1){0}}\put(30,95){\vector(0,1){0}}
\put(55,10){\vector(0,1){0}}
\put(30,53){\vector(0,1){0}}\put(55,53){\vector(0,1){0}}
\put(27,15){\framebox(31,28){\(P\)}}
\put(2,58){\framebox(31,28){\(Q\)}}
\put(10,7){\makebox(0,0){\(1\)}}
\put(35,7){\makebox(0,0){\(2\)}}
\put(60,7){\makebox(0,0){\(3\)}}
\end{picture}
\end{center}
\caption{Successive projection.\label{succproj}}
\end{figure}
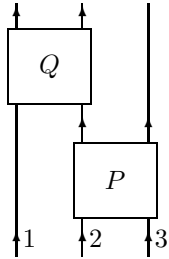

We'll treat just one special case for illustrative purpose. Consider the diagram in Fig. \ref{succproj}
where \(P\) and \(Q\) are rank-one projections, say onto vectors \(\Phi \in \cH_1\otimes \cH_2\) and \(\Psi \in \cH_2\otimes \cH_3\) respectively. Besides being a diagram of the type we're considering it's also meant to be
a {\em temporal\/} diagram meaning that time runs upward. Coecke associates a anti-linear map to each state, \(F_\Phi :\cH_2\to \cH_3\) and \(F_\Psi:\cH_1\to \cH_2 \) which can be defined by tensor universality for the case of product vectors by \(F_{\alpha \otimes \beta }:\phi \mapsto (\phi ,\alpha )\beta \). The theorem now states that if the state is of the form \(\phi _1\otimes \phi _{23}\in \cH_1\otimes(\cH_2\otimes\cH_3)\) then the output state is of the form \(\psi _{12}\otimes \psi _3\in (\cH_1\otimes\cH_2)\otimes\cH_3\)  where
\begin{equation}\label{revprocess}
\psi_3=F_{\Psi }\circ F_{\Phi }(\phi _1).
\end{equation}
The curious thing about this result is that the processing order implied by (\ref{revprocess}) is {\em opposite\/} to the temporal order, the later projection processes \(\phi _1\) first. This is a general feature of certain types of categories of which finite-dimensional Hilbert spaces is an example. This fact can be easily seen as the node of a rank-one projection splits into two nodes as shown in Fig. \ref{dsuccproj}.
\newsavebox{\coecke}
\savebox{\coecke}{
\put(5,0){\line(0,1){58}}\put(5,86){\line(0,1){10}}
\put(30,0){\line(0,1){15}}
\put(30,86){\line(0,1){10}}
\put(55,0){\line(0,1){15}}\put(55,43){\line(0,1){53}}
\put(5,10){\vector(0,1){0}}\put(30,10){\vector(0,1){0}}
\put(5,95){\vector(0,1){0}}\put(30,95){\vector(0,1){0}}
\put(55,10){\vector(0,1){0}}
\put(55,53){\vector(0,1){0}}
\put(27,15){\framebox(31,12){$\bar \Psi $}}\put(27,31){\framebox(31,12){\(\Psi \)}}
\put(2,58){\framebox(31,12){\(\bar \Phi \)}}\put(2,74){\framebox(31,12){$\Phi $}}
}
\begin{figure}[h!]
\begin{center}
\begin{picture}(60,96)
\usebox{\coecke}
\put(30,43){\line(0,1){15}}
\put(30,53){\vector(0,1){0}}
\end{picture}
\end{center}
\caption{Using rank-one property in Fig. \ref{succproj}\label{dsuccproj}}
\end{figure}
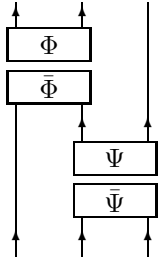

We now take the liberty of changing one line to its opposite to get the diagram:
\begin{center}
\begin{picture}(60,96)
\usebox{\coecke}
\qbezier[8](30,43)(30,50)(30,58)
\put(30,49){\vector(0,-1){0}}
\end{picture}
\end{center}
This is equivalent to the previous diagram and we see \(\bar\Phi \) and \(\Psi \) playing the roles of maps\footnote{These are not Coecke's maps being linear and connecting a \physical\ space to a \virtual\ one and vice-versa.} (state processors) acting, in fact, contrary to the temporal order. A diagrammatic treatment of Coecke's theorem and generalizations was given in \cite{svet:quant-ph/0601093}, however the present paper completely supersedes that one. Coecke's original proof used  fixed bases and a combinatorial induction, the one in \cite{svet:quant-ph/0601093} simplified this using tensor universality. The morphic approach makes the proof trivial. In morphic terms the diagram of Fig. \ref{dsuccproj} is expressed by \(\Phi ^{ab}\bar \Phi _{cd} \Psi ^{df}\bar \Psi _{gh}\). A little rewriting results in \(\{\Phi ^{ab}\bar \Psi _{gh}\}\{\bar \Phi _c{}^{\bar d} \Psi_{\bar d}{}^f\}\) and Coecke's theorem\footnote{Using linear instead of anti-linear maps.} is an obvious consequence. For this case of two projections, a proof using Dirac notation is as direct as this one, but for a general diagram with rank-one bipartite projection this is far from the case. In contrast a general morphic ``rewriting'' proof is almost as immediate as the one for two projections.

\newpage
\section{A Tall Tangled Tale\protect\footnote{Even annotated.} with Alice, Bob, Charlie, Diedre, Eve and a Quarrelsome Russian Sorceress.}\label{tale}

Once upon a time Ambitious Alice wanted to send Boyfriend Bob and unknown qbit, just like in the picture.
\newsavebox{\ttale}
\savebox{\ttale}{\put(0,75){\includegraphics[bb=0 0 20  30, scale=0.3]{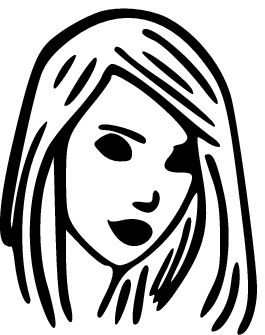}}
\put(125,125){\includegraphics[bb=0 0 20 30,scale=0.3]{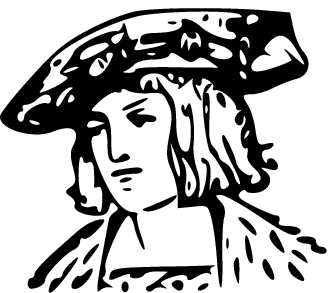}}
}
\begin{center}
\begin{equation*}\label{Alice2Bob}
\hbox{\begin{picture}(150,100)
\put(0,0){\includegraphics[bb=0 0 20  30, scale=0.3]{Alice.eps}}
\put(125,75){\includegraphics[bb=0 0 20 30,scale=0.3]{Bob.eps}}
\qbezier(30,20)(60,50)(90,50)
\qbezier(90,50)(120,50)(130,70)
\put(90,50){\vector(1,0){0}}
\end{picture}}
\end{equation*}
\end{center}
Sadly there was no direct quantum channel from Alice to Bob and she was stymied for a while until she remembered there was a channel from her to Charlie and another from Charlie to Bob. Alas, as a result of galactic global warming, Charlie was snared last weirdly warm winter by an Eight-headed, Eight-tailed Rogue Heterotic String and dragged off toward a hot event horizon where uncountable other such beasts were swarming.\footnote{Alice and Bob did plan to rescue him, but as from their perspective it would take infinite time for Charlie to cross the horizon, they were in no hurry. Also, they could make no plans without Evonymous Eve somehow learning bout it.} No help from there. Capricious Charlie did have the habit  of sending qbits to both of them which both stored in their own quantum memories, and though neither knew what it was all about, they dutifully kept them hoping one day to put them to some use. Alice, who was never one to pay much attention to physical laws, was then hit by a bright idea. She would send her qbit back in time to Charlie who would then send it on to Bob, just like in the picture.
\begin{center}
\begin{equation*}
\begin{picture}(150,150)
\usebox{\ttale}
\put(75,0){\includegraphics[bb=0 0 20 30,scale=0.3]{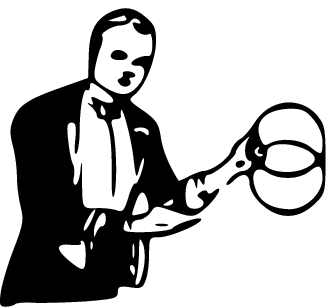}}
\qbezier(30,90)(40,100)(50,100)
\qbezier(50,100)(80,100)(80,60)
\qbezier(80,60)(80,40)(100,40)
\qbezier(100,40)(120,40)(130,120)
\put(80,60){\vector(0,-1){0}}
\end{picture}
\end{equation*}
\end{center}
Woefully, nothing in Charlie's laboratory notes, which Alice dutifully gathered up after his abduction, even remotely hinted at him having received anything from the future. She was sure her temporal inversions were working, but what was going wrong? Feeling a bit despondent, Alice sat brooding. Her thoughts ended up drifting to her friend Diedre's diagrams which showed temporal flows going ever which way and so ended up enclosing  each temporal turn in an \(I\)-box\footnote{Those that perceive  flaws and  contradictions in this fairy fable, here and hence, are politely asked to keep quiet.} as Diedre would have it, just like in the picture:
\begin{figure}[h!]
\begin{center}
\begin{picture}(150,150)
\usebox{\ttale}
\put(75,0){\includegraphics[bb=0 0 20 30,scale=0.3]{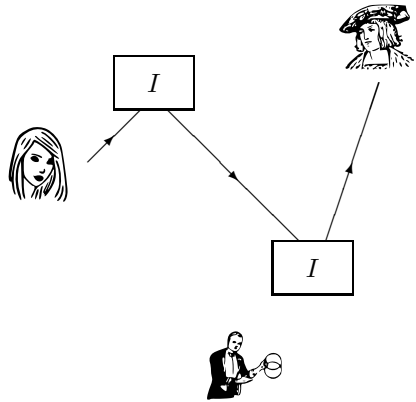}}
\put(40,110){\framebox(30,20){\(I\)}}
\put(100,40){\framebox(30,20){\(I\)}}
\put(30,90){\line(1,1){20}}
\put(60,110){\line(1,-1){50}}
\put(120,60){\line(1,3){20}}
\put(40,100){\vector(1,1){0}}
\put(87,83){\vector(1,-1){0}}
\put(130,90){\vector(1,3){0}}
\end{picture}
\end{center}\caption{Diedre's \(I\)'s}\label{AliceDiedre}
\end{figure}

\noindent Still no hint of progress. At last in a desperate attempt and not without much trepidation she called upon Baba Yaga\footnote{This is a fabulous character of Slavic fables (whereby she's a fabulous fabulous character). In olden days she lived in the forest in a little house on chicken's legs (much like the urban police booths one sees here and there) surrounded  by a fence made of human bones. She flew about in a mortar using the pestel as an oar. In modern times she's taken to urban living, gave up her ugly appearance, and drives an SUV run on biofuel (twisted tongues  say made from human humeri and female femurs) causing innumerable traffic jams and spreading road rage. She is often taken for an ordinary wicked witch, but this is an enormous error and a monstrous mistake. Her moral system is truly alien and meeting her can bring you either fortuitous fortune or ruinous ruin. By the way, her name is accented thus: B\'aba Yag\'a} for help. This Quarrelsomw Russian Sorceress was highly amused by the quandary and being in a good mood did something, as is her wont, unexpected and seemingly totally beside the point. She reversed the back-in-time q-flow between the two boxes to flow forward, just like in the picture:\
\begin{center}
\begin{equation*}
\begin{picture}(150,150)
\usebox{\ttale}
\put(75,0){\includegraphics[bb=0 0 20 30,scale=0.3]{Charlie.eps}}
\put(35,30){\includegraphics[bb=0 0 20 30,scale=0.3
]{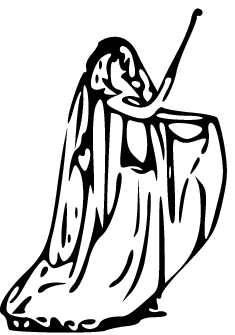}}
\put(40,110){\framebox(30,20){\(I\)}}
\put(100,40){\framebox(30,20){\(I\)}}
\put(30,90){\line(1,1){20}}
\put(60,110){\line(1,-1){50}}
\put(120,60){\line(1,3){20}}
\put(40,100){\vector(1,1){0}}
\put(77,93){\vector(-1,1){0}}
\put(130,90){\vector(1,3){0}}
\end{picture}
\end{equation*}
\end{center}
Due to the incompatibility with the \(I\)-nature of the boxes, this caused so much q-compression in the upper box and so much q-tension in the lower that the flow ruptured in both places just like in the picture:
\begin{center}
\begin{equation*}
\begin{picture}(150,150)
\usebox{\ttale}
\put(75,0){\includegraphics[bb=0 0 20 30,scale=0.3]{Charlie.eps}}
\put(40,110){\framebox(30,20){}}
\put(100,40){\framebox(30,20){}}
\put(30,90){\line(1,1){20}}
\put(60,110){\line(1,-1){50}}
\put(120,60){\line(1,3){20}}
\put(40,100){\vector(1,1){0}}
\put(77,93){\vector(-1,1){0}}
\put(130,90){\vector(1,3){0}}
\put(50,110){\line(0,1){40}}
\put(60,110){\line(0,1){40}}
\put(110,20){\line(0,1){40}}
\put(120,20){\line(0,1){40}}
\put(50,140){\vector(0,1){0}}
\put(60,140){\vector(0,1){0}}
\put(110,30){\vector(0,1){0}}
\put(120,30){\vector(0,1){0}}
\end{picture}
\end{equation*}
\end{center}
Alice was aghast! Her beautifully  planned temporal experiment ruined! {\it Why did I ever call upon that witch?!\/} she sobbed. She sat, her palms in her face, and cried. Suddenly among grins, snickers and chortles, she realized what had happened. Of course! The lower box was precisely Charlie's way of sending qbits to her and Bob. Charlie  simply suffered from the reinterpretation syndrome that ran rife among the tachyon traders. He thought the qbit he {\em received\/} from the future {\em from her\/} was mistakenly one  he {\em sent to her\/}  in the future. No wonder his notes said nothing. And the upper box? She asked Diedre who was conversant with such boxes. ``It's a measurement of course! It's like the lower box! How do you think Charlie got his entangled qbits in the first place?'' Diedre retorted somewhat disparagingly,  wondering why Alice didn't see that. But still, Bob did not get his qbit. As  euphoria wore off and depression threatened Alice  sent Bob (by ordinary e-mail, {\it What a letdown!}) a full account. {\it Whatever that may be worth!} She sat around in a blue funk and brooded.  Far away, Bright Boisterous Bob cried out ``Gotcha! Gotcha! Gotcha! Oh mysterious little qbit, I've got you now!" and WhatsApp'ed Alice the wonderful news. And so they lived happily ever after\footnote{One has to say that, even if you know it's not true, it's just a conventional fairy fable fabulation.} sending qbits back and forth and finally found a way to have private lover's chats without Eve listening in, just like in the picture:
\begin{center}
\begin{equation*}
\begin{picture}(150,200)
\put(0,75){\includegraphics[bb=0 0 20  30, scale=0.3]{Alice.eps}}
\put(125,175){\includegraphics[bb=0 0 20 30,scale=0.3]{Bob.eps}}
\put(75,0){\includegraphics[bb=0 0 20 30,scale=0.3]{Charlie.eps}}
\put(40,110){\framebox(30,20){}}
\put(100,40){\framebox(30,20){}}
\put(30,90){\line(1,1){20}}
\put(60,110){\line(1,-1){50}}
\put(120,60){\line(1,5){23}}
\put(40,100){\vector(1,1){0}}
\put(77,93){\vector(-1,1){0}}
\put(132,120){\vector(1,4){0}}
\put(50,110){\line(0,1){90}}
\put(60,110){\line(0,1){90}}
\put(110,20){\line(0,1){40}}
\put(120,20){\line(0,1){40}}
\put(50,170){\vector(0,1){0}}
\put(60,170){\vector(0,1){0}}
\put(110,30){\vector(0,1){0}}
\put(120,30){\vector(0,1){0}}
\qbezier[40](50,130)(90,150)(130,170)
\qbezier[40](60,130)(100,150)(140,170)
\put(90,145){\vector(2,1){0}}
\put(90,150){\vector(2,1){0}}
\end{picture}
\end{equation*}
\end{center}
And so Patient People, that's how, thanks to Baba Yaga, quantum teleportation was born. But the story doesn't end here\ldots

Diligent Diedre was very excited by the development and went about drawing diagrams and doing calculation. She had trouble though dealing in an elegant way with all the e-mails and WhatsApp exchanged between Alice and Bob that were needed for teleportation to work. She kept mulling about this until one day her Muse sent her an idea and so one late evening she went to Alice's q-lab and said:

{\em Diedre} -- Alice, {\sl Amica\/}, your teleportation scheme is wonderfully interesting, exciting and mysterious.

{\em Alice} -- Thank you, Diedre Dear, Bob and I are now gearing up to do qtrits which will be really great! How are your calculations coming along?

{\em Diedre}  -- Well, that's why I'm here. I've had trouble dealing with all that CC. I never understood what that meant but then realized it means Carbon Copies of those two qbits that result from your projective measurement. Well, those two qbits carry exactly the same information that the two cbits of CC do, so if you send those two qbits to Bob all he need do is feed them along with his part of the entangled pair into a proper unitary, just like in the diagram
\begin{center}
\begin{equation*}
\begin{picture}(90,190)
\put(0,70){\framebox(30,20){\(M\)}}
\put(60,0){\framebox(30,20){\(\Omega \)}}
\put(70,20){\line(-1,1){50}}
\put(10,0){\line(0,1){70}}
\put(80,20){\line(0,1){120}}
\put(10,90){\line(1,1){50}}
\put(20,90){\line(1,1){50}}
\put(50,140){\framebox(40,20){\(U\)}}
\put(60,160){\line(0,1){30}}
\put(70,160){\line(0,1){30}}
\put(80,160){\line(0,1){30}}
\end{picture}
\end{equation*}
\end{center}
and then one doesn't need CC at all and the math, I'll bet, is simpler. This is a great idea and I'm sure Hugh\footnote{Diedre, time and again, speaks of Hugh though nobody knows who he is. Could he be an old boyfriend? It does seem at times that Diedre hales from another universe!} would have loved it.

{\em Alice} -- Diedre, you're daft! If I could send qbits directly to Bob I would not have needed the teleportation scheme to begin with and would not have had to risk my soul with that Rascally Russian Rusalka.\footnote{Baba Yaga is not a Rusalka, but Alice could never keep these kinds of  beings straight.}

{\em Diedre}  -- But Alice, the math calls for it and \ldots

{\em Alice} -- You mathematicians are so infuriating! You're only interested in  your Cute Calculations and what's ``Obvious and Elegant". Only to yourselves of course! Such Categorical Frivolity! Back in the Real World, I and Bob are going to be powerful QDUDES, are about to create a huge Q-FIRM, and will put Microsoft, Google, and Facebook out of business, besides \ldots

{\em Diedre}  -- Ok, Ok, Alice, cool it! I thought you might be interested. I guess I'll just shut up and calculate.

Alice quieted down and Diedre began  drawing fervently like one possessed for she finally saw how her latest diagram, which she thought was just symbolic, was actually right on the nose. {\sl Oh Hugh, Hugh,}-- she thought-- {\sl you wouldn't have liked this, but as you well know, math speaks loudest of all.}

And that's how, Doubly Patient People, thanks to Alice, the SHUAC\footnote{To get away from all those capital letters, I will from now on write ``shuac", which is SHUAC lite.} approach to quantum mechanics was born. But the story does not end here\ldots

As you all must suspect, that eventful eventide Eve
\vskip 20pt
\begin{center}
\includegraphics[bb = 0 20 30 20, scale=0.3]{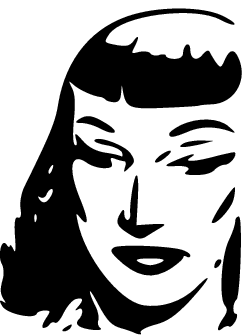}
\end{center}
 was eavesdropping, and even as evening fell and she crept to her hideout, passing under the eaves of an ancient abandoned post-office, a site of  evil rituals of times long past, she was struggling with an evanescent thought which finally, harking back to her failed career as a High-Energy Particle Prophetess,\footnote{Her would-be thesis, ``Everything is held together by neutrinos", was demolished by the Gauge Gang.} she expressed, taking a hint from Diedre's diagram, as a reaction:\footnote{Eve didn't quite get it right. But then, fairy tales are not scientific texts, their characters are not science-literate, their conclusions are not reality-checked, and their hypotheses are not fully formulated.}
\begin{equation}\label{evesdream}
\hbox{Back-in-time q-flow}\rightarrow 3 \times \hbox{Forward-in-time q-flow}.
\end{equation}
Struggling to understand this she wondered if there was a reverse reaction by which she could eventually build a time machine, go back in time, and stop herself from pledging allegiance to  Baba Yaga.  But the story doesn't end even here\ldots

\section{What's Real? What's Local? What's Space? What's Time?}\label{whatswhat}

Not every propic diagram corresponds to a physically realizable play in the laboratory. There are several laboratory interpretations that can be given to a diagram, here we deal with one that is closest to those given to similar diagrams appearing in the literature. The least requirements should be:
\begin{enumerate}
\item The diagram has to be {\em temporal\/} in which processes are placed in temporal order on the page and we usually take time to increase from bottom to top.
\item At any time the existing lines represent a possibly entangled n-partite state, with each line representing each separate part.
\item Given the previous item, each line must be \physical\ and not \virtual\ and each must always point in the same {\em upward\/} direction.
\item The lines at any time must all represent {\em different\/} Hilbert spaces.
\item The nodes must correspond to physically realizable quantum situations, which at the point we are now can only be unitary transformations, state preparations or measurements. A state is represented by a node with only outgoing \physical\ lines and measurements one with only incoming \physical\ lines. These do not fully describe state preparation or measurement as one is not representing those situations in which the preparation procedure fails or when the measurement result corresponds to a projection onto a state different from the one in the diagram. Also generalized measurements with POVMs are not represented. \footnote{The construct in section \ref{woods} can be made to handle this, but we do not explore this aspect.}
\end{enumerate}

\begin{figure}[h!]
\begin{center}
\begin{picture}(90,160)
\put(0,70){\framebox(20,20){\(U\)}}
\put(10,90){\line(1,1){50}}
\put(35,115){\vector(1,1){0}}
\put(50,0){\framebox(30,20){\(\Omega \)}}
\put(60,20){\line(-1,1){50}}
\put(30,50){\vector(-1,1){0}}
\put(70,20){\line(0,1){120}}
\put(70,80){\vector(0,1){0}}
\put(65,150){\makebox(0,0){Bob}}
\put(30,10){\makebox(0,0){Charlie}}
\put(35,80){\makebox(0,0){Alice}}
\end{picture}
\end{center}
\caption{Superdense coding\label{sdc}}
\end{figure}
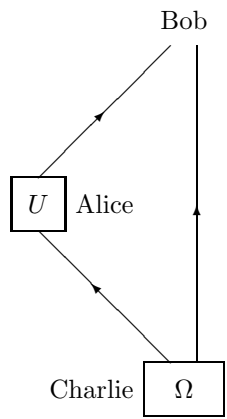

Though there may be other conditions that one should impose, this for now is enough to proceed. What is truly amazing is that a proper physically realizable diagram can be converted to a mathematically equivalent one seemingly having no direct physical interpretation. The morphs at the nodes can be made to play different roles, simply by changing lines to their opposites. A case in point is the superdense coding scheme for qbits as in Figure \ref{sdc}.
Here Charlie is a source of maximally entangled qbits (one of the Bell states) and Alice has a choice of applying a unitary to her part. After doing so, she sends the resulting qbit to Bob. By a proper choice among four unitaries she succeeds that Bob can receive from the two sources four orthogonal two-qbit states, and so she succeeds in sending two cbits of information with each qbit, hence the moniker superdense. Now there is a famous bound on the ammount of information one can send via a {\em direct\/} quantum channel, that is, not using any shared resource such as an entangled state. This bound is known as the Holevo bound and it states, without going into precise numerics, that for each cbit one has to send at least one qbit. Hence it is often stated that superdense coding violates the Holevo bound and many have found this as part of the ``quantum mysteries" offered up by entanglement. The Horodeckis\cite{horofam:RMP81.865} offer the following remarks toward a possible reconciliation:
\begin{quote}
Why does not this contradict the Holevo bound? This is because the
communicated qubit was a priori {\it entangled} with Bob's qubit. This
case is not covered by Holevo bound, leaving place for this strange
phenomenon. Note also that as a whole, two qubits have been sent: one
was needed to share the EPR state. One can also interpret this in the
following way: sending first half of singlet state (say it is during
the night, when the channel is cheaper) corresponds to sending one bit
of {\it potential communication}. It is thus just as {\it creating the
  possibility} of communicating $1$ bit in future: at this time Alice
may not know what she will say to Bob in the future. During day, she
knows what to say, but can send only one qubit (the channel is
expensive). That is, she sent only one bit of {\it actual
  communication}. However sending the second half of singlet as in
dense coding protocol she uses both bits: the {\it actual} one and
{\it potential} one, to communicate in total $2$ classical bits. Such
an explanation assumes that Alice and Bob have a good quantum memory
for storing EPR states, which is still out of reach of current
technology. In the original dense coding protocol, Alice and Bob share
pure maximally entangled state.
\end{quote}

Then there is the view attributed to Schumacher in \cite{benn-wies:PRL69.2881}
\begin{quote}
It therefore might be better to say, as Schumacher suggests,\footnote{The original had a reference number here whose content was ``private communication"} that one of the two bits is sent forward in time through the treated particle, while the other bit is sent backward in time to the EPR source, then forward in time through the untreated particle, until  finally it is combined with the bit in the treated particle to reconstitute the two-bit message.
\end{quote}

Now if for the moment we suspend any criteria of physical reality and think of the two remarks as fairy tales, which one is more interesting and easier to follow? Which one would you translate into kid talk and tell your children?

Figure \ref{sdc} is a propic diagram and as such there is another mathematically equivalent one given in Figure \ref{sdcgood}:
\begin{figure}[h!]
\begin{center}
\begin{picture}(120,40)
\put(0,0){\framebox(20,40){\(U\)}}
\put(20,30){\line(1,0){100}}
\put(70,30){\vector(1,0){0}}
\qbezier[15](20,10)(35,10)(50,10)
\put(35,10){\vector(1,0){0}}
\put(50,0){\framebox(20,20){\(\Omega \)}}
\put(70,10){\line(1,0){50}}
\put(100,10){\vector(1,0){0}}
\end{picture}
\end{center}
\caption{Superdense coding, channel version\label{sdcgood}}
\end{figure}
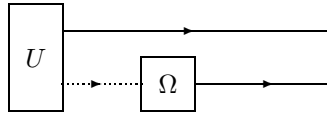

Here, of course, the dotted line is a ``co-qbit going backward in time", if so one wishes to think. This diagram is in ``channel style", that is, successive processes take place from left to right, and so ``time" runs from left to right. This is the normal convention in communication theory.

Now to a shuac mathematician a Hilbert space is a Hilbert space is a Hilbert space, and a channel is a channel is a channel. A \virtual\ Hilbert space is as much a Hilbert space as a \physical\ one. The above channel diagram is precisely one for which the Holevo bound holds. Alice has as her disposal an alphabet of ``states" of the form \(\sum u_{ij}e_i\otimes \bar e_j\) where \(u_{ij}\) is a unitary matrix and \(e_i\) a basis for qbits. In morph notation her states are \(U^{\bar ab}\), unitary transformations in another role. Charlie's state creation \(\Omega \) takes on the role of a linear transformation, \(\Omega _{\bar a}{}^b\) in morph notation, and doing the math one sees that for this channel the Holevo bound is respected.
{\em The Holevo bound in quantum superdense coding is rigorously maintained once one interprets the channel properly.}
One sees thus that Schumacher's remark is right on the money as far as counting correctly is concerned.

One may wonder how a shared resource can become a channel. There is nothing ``quantum'' about this, it can be done in classical communication. If Alice and Bob have shared knowledge then sending a single cbit can convey a world of information, for instance a one cbit {\em yes\/} in the context of (shared) previous talks of marriage and moving far away from Eve's interference. More to the point, as the {\em meaning\/} of a message is of no concern in information theory, one can use time as a shared resource. Thus in a more prosaic story if both Alice and Bob have perfectly synchronized clocks and the  time interval between sending and receiving can be rigorously controlled, one can divide a time period, of say a second, into, say sixteen subintervals, and then if Alice sends one ``1'' cbit, Bob can determine in which subinterval the message originated and so associate it to four cbits. See \cite{burc:arXiv:0711.0356} for another scheme for using time as a ``channel". In \cite{coll-pope:PRA65.032321} it is argued that ``secret communication'' is a classical channel analogous to shared entanglement. To a shuac mathematician, ``secret'' is beside the point. Public communication to which no one listens or listens but doesn't care (like that sourced by many of our politicians) would do just as well. Many classical situations can be forced to display properties claimed to be typically quantum, but this helps very little, if at all, in understanding what being quantum is really all about, it just shows that we've neglected some interesting classical constructs.

If we complete Figure \ref{sdcgood} by a specific state of Bob's measurement basis, we get a diagram with three nodes and no outgoing or incoming lines:
\begin{center}
\begin{picture}(120,40)
\put(0,0){\framebox(20,40){\(U\)}}
\put(20,30){\line(1,0){80}}
\put(65,30){\vector(1,0){0}}
\qbezier[15](20,10)(35,10)(50,10)
\put(35,10){\vector(1,0){0}}
\put(50,0){\framebox(20,20){\(\Omega \)}}
\put(70,10){\line(1,0){30}}
\put(90,10){\vector(1,0){0}}
\put(100,0){\framebox(20,40){\(\Lambda \)}}
\end{picture}
\end{center}
Since we can change the direction of any line by changing it to the opposite one, this diagram can be changed to be a channel going from any node to any other, six channels in total. They may not all be relevant to the original quantum dense coding problem, but do illustrate the multiple ways a propic diagram can be interpreted. In classical communication such alternate channels are not discussed but once again it's more a question of neglect than lack of ``quantumness". The simplest version of classical channel is just a map \(T:X\to Y\)  between two sets. Applying any contravariant functor \(F\), such as \(\Hom(\cdot,Z)\) produces a map \(F(T):F(Y)\to F(X)\) which can be called a ``dual channel". Such a channel is not usually operative, in the sense that no message is transmitted through it while the direct channel is operating, but it has it's manifestations. Alice calls Bob on her cell phone and tells him a story. They meet later and Alice, to her dismay, finds out that Bob didn't really get it right, twisting everything she said to conform to his idiosyncratic view of things. This is the dual channel functioning where \(F\) is a map from spoken words to mental states. In the quantum prop, line reversals can be done without introducing contravariant functors, and the other channels are more apparent.

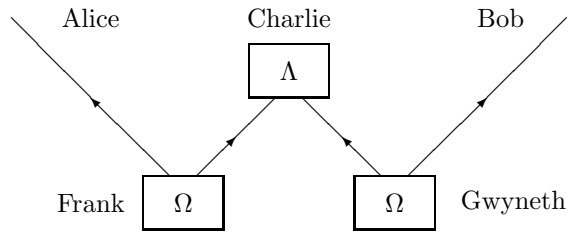
\begin{figure}[h!]
\begin{center}
\begin{picture}(210,85)
\put(50,0){\framebox(30,20){\(\Omega \)}}
\put(130,0){\framebox(30,20){\(\Omega \)}}
\put(90,50){\framebox(30,20){\(\Lambda \)}}
\put(70,20){\line(1,1){30}}
\put(85,35){\vector(1,1){0}}
\put(140,20){\line(-1,1){30}}
\put(125,35){\vector(-1,1){0}}
\put(60,20){\line(-1,1){60}}
\put(30,50){\vector(-1,1){0}}
\put(150,20){\line(1,1){60}}
\put(180,50){\vector(1,1){0}}
\put(185,80){\makebox(0,0){Bob}}
\put(105,80){\makebox(0,0){Charlie}}
\put(30,80){\makebox(0,0){Alice}}
\put(30,10){\makebox(0,0){Frank}}
\put(190,10){\makebox(0,0){Gwyneth}}
\end{picture}
\end{center}
\caption{Entanglement swapping.\label{entswap}}
\end{figure}

Another instructive situation occurs in {\em entaglement swapping\/} which we display in Figure \ref{entswap}, again just for qbits.
Here Frank and Gwyneth are sources of the same maximally entangled state, a Bell state. Charlie performs a projective measurement in the Bell basis, and \(\Lambda \) is one of these basis states. One discovers now that the two-qbit state held by Alice and Bob is entangled. This to some seems mysterious for how can Charlie's actions, which are far removed from both Alice and Bob, entangle two qbits that started out not entangled and  always remained widely separated. There is a catch though. Charlie has to inform Alice and Bob just for which pairs of qbits his measurements resulted in a projetion onto \(\Lambda \). The total two-qbit ensemble held by Alice and Bob corresponding to  {\sl all\/} of Charlie's measurement results is not entangled, but the subensemble  corresponding to the \(\Lambda \) result is.\footnote{After Alice and Bob became Riotously Rich and Whoopingly Wealthy, they  traveled to the opposite edges of the universe (How? Wise women will whisper ``warp'') capture a herd of Wild Wilson Loops and harvest a field of {\sl Praecursor potens\/} with which they extricated Charlie from the swarm of Heterotic Strings using entanglement swapping and some quantum tricks that  are carefully guarded trade secrets (such as how they  communicated with Charlie -- wary warriors will whisper ``wormholes"). They made Charlie the CFO of their q-firm where he developed q-money that can be spent and saved at the same time, getting around the famous Superselectman's ruling forbidding such double actions by a subtle loophole that was discovered by Bob's legal acumen. They  lived happily ever after and the story  ends here. This is a fairy tale made up to teach quantum mechanics to children. The true story is other\ldots } Nevertheless, it would be hard to argue that such communication would {\em create\/} the entanglement and the mystery, to many, still remains.

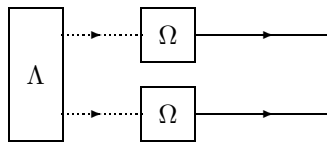
\begin{figure}[h!]
\begin{center}
\begin{picture}(120,50)
\put(0,0){\framebox(20,50){\(\Lambda \)}}
\qbezier[15](20,10)(35,10)(50,10)
\qbezier[15](20,40)(35,40)(50,40)
\put(35,10){\vector(1,0){0}}
\put(35,40){\vector(1,0){0}}
\put(50,0){\framebox(20,20){\(\Omega  \)}}
\put(50,30){\framebox(20,20){\(\Omega  \)}}
\put(70,10){\line(1,0){50}}
\put(100,10){\vector(1,0){0}}
\put(70,40){\line(1,0){50}}
\put(100,40){\vector(1,0){0}}
\end{picture}
\end{center}
\caption{Entanglement swapping, channel version.\label{entswapgood}}
\end{figure}

The channel version of entanglement swapping is given by Fig. \ref{entswapgood}.
 In this reading of the play, Charlie is a {\em source\/} of the entangled Bell state \(\Lambda \) and Frank and Gwyneth perform local invertible operations. It's obvious that Alice and Bob's two-qbit state is entangled. {\sl Entanglement swapping works precisely for the same reason that invertible local operations cannot destroy entanglement.} To a shuac mathematician, a local operation is a local operation is a local operation, and there's no mystery here.

Suppose one is given two explanations of a phenomenon. The first is conceptually clear and mathematically easy, the second is conceptually obscure and mathematically awkward. Which explanation would one choose to be the closest to physical reality? The first one of course, {\em except for quantum mechanics.} In quantum mechanics one discards conceptual clarity and easy mathematics  for conceptual obscurantism and difficult math. And for what reason exactly?
Is it time and causality? Figures \ref{sdc} and \ref{entswap} represent the real world, while Figures \ref{sdcgood} and \ref{entswapgood} a fictitious world. In the real world there are no funny ``co-states" (represented by \virtual\ lines) going backward in time and only unitary transformations are realizable. In the fictitious world there are the abberations, at least, of back-in-time flow and arbitrary linear transforms. Clearly the real world view is preferable. So the real world explanations are obscure and the math is hard while the fictitious world explanations are clear and the math is easy, but {\em the two are mathematically equivalent!} This is a most fascinating quandary we've gotten ourselves into. If math speaks loudest of all, something must be done. In the history of physics, when ontology conflicted with math,  math triumphed and ontology changed and conformed to the math. The luminiferous ether gave way to fields and space and time gave way to space-time under the sway of Maxwell's equations and Lorentz transformations, math in short. Is it time to perform some sort of ontological cleansing to extricate ourselves from the quantum quandary? Let us see how much we can get away with.

Surely we must maintain causality and banish back-in-time flows. Otherwise\ldots\ But listen, I\footnote{This may seem a break of character going over to first person singular in the text. But we remind the readers that in fairy tales all thing are opposite, and so not so.} overhear some chatter, why it's Eve and Bab Yaga, her mentor and, at times, her tormentor\ldots

{\em Eve} -- \ldots very clever the way you switched the direction of Alice's q-flow. You changed the past.

Eve, as we know, is obsessed with finding a way to build a time machine.

{\em Yaga} -- I just put things right, I {\em am\/} a force of Nature, you realize, and you can't change the past. You can {\em affect\/} the past, that's completely different,  easy to do, and I do it all the time.

{\em Eve} -- Isn't that the same thing? If you can act on the past, can't you then change it.

{\em Yaga} -- Not at all, what's done is done. You make something happen in the past, you can't then undo it. Do it once and that's it! For instance:

She picked up a hard-boiled painted egg, peeled it, salted it, and swallowed it whole. She then sat quietly and expectantly. After some time\ldots

{\em Eve}  -- For instance what? Weren't you going to show me something?

{\em Yaga} -- I did. By eating that egg I affected the past.

{\em Eve} -- I don't see it, affected what?

{\em Yaga} -- Why the very events that put that egg on my table. I get all my food this way. You don't see me shopping at Wall Mart do you?

Eve decided to be the devil's advocate. {\sl If I argue hard enough that she can't affect the past, maybe she'll slip up and tell me how in fact one can change it.}

{\em Eve} -- This is very confusing to me. You are saying that by eating your food you have caused it to come to you. I'll prove that you couldn't have done that. By causality\ldots

{\em Yaga} --  Eve, careful! If you invoke causality you are {\sl assuming \/} pretty much the conclusion you are trying to prove. That's circular. Remember when you went circular on me? [{\em Eve shuddered.\/}] You have to argue from all the physics and math you know setting causality aside. Causality should be the {\em conclusion\/}.  So?

{\em Eve} -- Ok, fair enough. You see there's a paradox if you\ldots

And try as she could, Eve could not find a strong enough argument. Her attempt was nipped in the bud and her heart sank for if she couldn't get hold of {\em some\/} Timely Time Tricks (she was sure Yaga had plenty) how will she ever extricate herself from the clutches of this Wizardly Witch? Somewhat halfheartedly she went on:

{\em Eve} -- But how do you know that what you're about to do is going to cause and event that's already happened.

{\em Yaga} -- Do you see sorceresses revealing their secrets?

{\em Eve}  -- And what if all of a sudden you decided not to be the cause, you do have free will, don't you? Then how is the event that already took place to have\ldots, to have been\ldots, to have been happened\ldots Ah, you know what I mean! I suppose though that if you only know of the cause-effect relation after the fact of both events, that little conundrum won't come up. What's cause and what's effect is then convention! Oh, I give up!

Eve sat dejected and Baba Yaga gave her an mysterious smile with just a hint of warmth in her placid gaze. It was one of those rare moments when somewhere in the abyss of her ancient alien soul she did feel a strange affection for her favorite spy. {\em You'll really be something else again my fine fey fledgling, when I finally set you free to fly!}

After the fact of two events it {\em is\/} conventional as to which is cause and which is effect if viewed in sufficient isolation. This may not work in a court of law. No defense lawyer would argue that by dying from a bullet, the victim actually {\em caused\/} the past event of the killer shooting him. If he did the jury would have reasonable doubts about his sanity, but the prosecution would not be able to prove him wrong if restricted by Baba Yaga's instruction. Laboratory plays have to be sufficiently isolated from various influences  to be effectively modeled by propic diagrams and causality within such a restricted context is something else again. What (\ref{caniso}) and the {\em mathematical\/} equivalence of Fig. \ref{sdc} and  Fig. \ref{sdcgood} and likewise of Fig. \ref{entswap} and \ref{entswapgood} really say, as they deal with alternate descriptions of the {\em same\/} physical reality, is that one has a {\em gauge freedom\/}, the gauge freedom to switch cause and effect in certain contexts. The gauge nature of this is generally not recognized and so, for instance, the report of Shumacher's remark in \cite{benn-wies:PRL69.2881} is followed by an assurance that the back-in-time flow to which Shumacher refers to cannot be used to violate causality. Well, obviously, no {\em gauge\/} transform of a causal description, such as Fig. \ref{sdcgood}, can be the basis of acausal physics.\footnote{Unless one contemplates  breaking the gauge symmetry.} Once the gauge nature of such cause and effect switching is realized, no such comments are necessary. A propic diagram is thus causal, in  a simple sense and as a first attempt at such a notion, not because all lines are \physical\ and lead to the future, but because it can be {\em transformed\/} to one such by the gauge freedom given by (\ref{caniso}). The switch from state to channel, such as Frank's and Gwyneth's arrangements in Figures \ref{entswap} and \ref{entswapgood} are also gauge transformations. That after such switches one can still think of the result in understandable terms ``channel'', ``back-in-time flow'' (weird but understandable), ``local operation",  is a boon for we can construct a mental picture of what is ``happening'' and make no excuses for the fact that this ``happening'' goes on in a fictitious world if this simplifies the story and the math. {\em Si non \'e vero, \'e ben trovato!\/}\footnote{Translating from Italian to folk American: ``If it ain't true, it's still a darn good yarn!''} as they say.

The forward-in-time vs.\ backward-in-time gauge freedom has been around for a long time in certain contexts without ever raising an eyebrow. In the Shr\"odinger picture\footnote{In most of this text we're implicitly working in the Heisenberg picture.} if we consider the inner product of a state \(\Psi \) with a time-evolved state \(U(t)\Phi \) then
\begin{equation}\label{inpgauge}
(\Psi ,U(t)\Phi )=(U(-t)\Psi ,\Phi ),
\end{equation}
which is quite familiar.
Thus the inner product of \(\Psi \) with the {\em forward-in-time\/} evolved \(\Phi \), is the same as the the inner product of the {\em backward-in-time\/} evolved \(\Psi \) with \(\Phi \).
This is only indirectly the gauge freedom we've been discussing. In propic diagrams, equation (\ref{inpgauge}) is best seen as
associativiy:
\begin{center}
\begin{picture}(110,110)
\put(5,5){\framebox(30,20){\(\Phi \)}}
\put(5,45){\framebox(30,20){\(U(t) \)}}
\put(5,85){\framebox(30,20){\(\bar\Psi \)}}
\put(20,25){\line(0,1){20}}
\put(20,65){\line(0,1){20}}
\put(20,40){\vector(0,1){0}}
\put(20,80){\vector(0,1){0}}
\put(75,5){\framebox(30,20){\(\Phi \)}}
\put(75,45){\framebox(30,20){\(U(t) \)}}
\put(75,85){\framebox(30,20){\(\bar\Psi \)}}
\put(90,25){\line(0,1){20}}
\put(90,65){\line(0,1){20}}
\put(90,40){\vector(0,1){0}}
\put(90,80){\vector(0,1){0}}
\put(55,55){\makebox(0,0){\(=\)}}
\put(0,0){\framebox(40,70)}
\put(70,40){\framebox(40,70)}
\end{picture}
\end{center}
This becomes apparent once one realizes that the node corresponding to the contour containing the two upper nodes on the right-hand side is the Riesz dual of \(U(-t)\Psi \). This can be seen from:
\begin{center}
\begin{picture}(150,85)
\put(0,25){\framebox(30,20){\(U(-t) \)}}
\put(0,65){\framebox(30,20){\(\Psi \)}}
\put(15,5){\line(0,1){20}}
\put(15,45){\line(0,1){20}}
\put(15,10){\vector(0,-1){0}}
\put(15,50){\vector(0,-1){0}}
\put(60,25){\framebox(30,20){\(\overline{U(-t)} \)}}
\put(60,65){\framebox(30,20){\(\bar\Psi \)}}
\put(75,10){\vector(0,-1){0}}
\put(75,50){\vector(0,-1){0}}
\put(120,25){\framebox(30,20){\(U(t) \)}}
\put(120,65){\framebox(30,20){\(\bar\Psi \)}}
\put(135,5){\line(0,1){20}}
\put(135,45){\line(0,1){20}}
\put(135,20){\vector(0,1){0}}
\put(135,60){\vector(0,1){0}}
\qbezier[10](75,5)(75,15)(75,25)
\qbezier[10](75,45)(75,55)(75,65)
\put(45,42){\makebox(0,0){\(\stackrel{\overline{(\cdot)}}\mapsto\)}}
\put(105,42){\makebox(0,0){\(=\)}}
\end{picture}
\end{center}
where on the left one has the transformation from a diagram to its Riesz dual, and the equality follows from the freedom to change lines to their opposites and the fact that for a unitary group \(\overline{U(-t)}\) is the morphic transpose of \(U(t)\) which then changing the lines to their opposites gives the rightmost diagram.

Dirac notation, in it's wonderful ambiguity, shows this associativity simply:
\[\bra\Psi{U(t)}\ket\Phi =(\bra\Psi U(t))\ket \Phi =\bra\Psi (U(t)\ket \Phi )\]

In textbooks on particle physics one still comes across the metaphor that antiparticles are ``particles traveling backward in time'' and Feynman diagrams, if drawn in space-time, often keep up this pretence. Once Feynman diagrams move to momentum space one forgets about this little bit of folklore, expanding momentum space solutions of the corresponding wave equations into positive and negative energy parts, assigning one to the particles and the other to the antiparticle.  Of course anti-particle states are not ``co-states'', and their alleged ``going backward in time'' is not strictly speaking the same as for the ``co-states'' in Figures \ref{sdcgood} and \ref{entswapgood} but the difference is due to the  PCT theorem. Let \(\Theta \) be the PCT operator on a physical Hilbert space \(\cH\), then it can be viewed as a {\em linear\/} map  \(\Theta :\cH^*\to H\). If now \(\Psi \in \cH\) is an electron state, ``going forward in time'', then it's Riesz transform \(\bar \Psi \) is, after changing the lines attached to its node to the opposites, the corresponding ``co-state going backward in time''. Finally \(\Theta \bar\Psi \) is a {\em positron\/} state once again ``going forward in time''. This makes relativistic quantum field theory somewhat oblivious to the gauge freedom of changing cause and effect. Diagrammatically  (with time running upward) what we've just said is:
\begin{center}
\begin{picture}(210,80)
\put(0,30){\framebox(30,20){\({\rm e}^- \)}}
\put(15,0){\line(0,1){30}}
\put(15,20){\vector(0,1){0}}
\put(15,50){\line(0,1){30}}
\put(15,70){\vector(0,1){0}}
\put(60,30){\framebox(30,20){\(\overline{{\rm e}^-} \)}}
\qbezier[10](75,0)(75,15)(75,30)
\put(75,20){\vector(0,1){0}}
\qbezier[10](75,50)(75,65)(75,80)
\put(75,70){\vector(0,1){0}}
\put(120,30){\framebox(30,20){\(\overline{{\rm e}^-} \)}}
\put(135,0){\line(0,1){30}}
\put(135,15){\vector(0,-1){0}}
\put(135,50){\line(0,1){30}}
\put(135,65){\vector(0,-1){0}}
\put(180,30){\framebox(30,20){\({\rm e}^+ \)}}
\put(195,0){\line(0,1){30}}
\put(195,20){\vector(0,1){0}}
\put(195,50){\line(0,1){30}}
\put(195,70){\vector(0,1){0}}
\put(45,40){\makebox(0,0){\(\stackrel{\overline{(\cdot)}}\mapsto\)}}
\put(105,40){\makebox(0,0){\(=\)}}
\put(165,40){\makebox(0,0){\(\stackrel\Theta\mapsto\)}}
\end{picture}
\end{center}
The nodes of this diagram are ``propagators'', that is, time evolution operators for some fixed period of time. This discussion also explains why time-reversal operations in quantum mechanics are generally anti-linear. To reverse the time flow one has to go to the Riesz dual and to get back to the \physical\ Hilbert space one applies a {\em linear\/} operator from the \virtual\ space to the \physical\ one. The whole procedure is anti-linear.\footnote{Some systems with Hamiltonians whose spectrum is symmetric under reflection in \(0\) can be time-reversed by a linear operator, but this is very special.}

This whole causality question is of course tied up with the ``arrow of time'' problem. What makes the quantum mechanical situation described by propic diagrams different is that one can reverse the time direction of any line alone, changing it to its opposite. One thus has {\em local\/} time reversal symmetry of sorts.  This situation does not seem to have an obvious classical counterpart.\footnote{In principle one should be able to construct a classical counterpart to any ``quantum feature'', if for no other reason than the fact that quantum mechanics can be viewed as restricted classical mechanics, or that one can simulate quantum mechanics on classical computers, or that quantum mechanics is formalized by mathematics whose roots are classical (measuring land, counting sheep, etc.). Just how natural or instructive such classical counterparts are is a separate issue. It seems they don't provide any true insights into quantum mechanics, after all they {\sl are\/} ``just classical".}

Now one can argue that ``co-states'' are not real states and what is needed is some {\em global\/} principle stating that there has to be a gauge in which all Hilbert space lines are \physical\ and upward leading and this would define an overall arrow of time shared by all states. The situation is however more complex than this.\footnote{From the shuac perspective, dual Hilbert spaces are just as ``Hilbert" as any other so the proposed principle is too simplistic.}

There are two remaining issues. One concerns \physical\ lines going to the past, as in Alice's time reversals in Fig. \ref{AliceDiedre}, and the other concerns the legitimacy of Baba Yaga's injunction that causality has to be emergent. The first issue cannot be dealt with further without additional  mathematical development and we shall return to it at times when more can be said. In relation to Baba Yaga, emergent causality has been a desideratum  of any number of ``fundamental'' theories which nowadays usually fall under the moniker ``quantum gravity''. To have space-time along with its causal structure emerge is notoriously difficult and no one has succeeded to general content. Some approaches such as Sorkin's causal sets\cite{sork} or Loll's
causal triangulation\cite{loll-who} actually {\em put causality in by hand at the beginning\/}, as Eve would have had it,  and this seems to be somehow more successful at first glance. In view of the gauge freedom discussed above this might mean that ``quantum gravity'' is not quite as ``quantum'' as the rest of quantum mechanics, but the jury is still out and all one can say at this point is that it's too soon to decide between emergent and built-in causality.

We have succeeded nevertheless in a small ontological clean-out, cause and effect and the description of quantum processes are not as rigid as they once may have seemed. It's time to probe another aspect, the lines representing ``states'', especially ``entangled'' ones, themselves. Once more we're interrupted by chatter. We've gone back in time to when Eve was a graduate student attending a quantum mechanics course by Joana von Alteweib, a q-{\sl femme fatale\/} of far-flung fame

{\em von} -- \ldots\ and as you can see, in an entangled two-part state, the parts don't have a state of their own.

{\em Eve} -- But Professor, each part is an entity existing in the world, and it exist in some way that distinguishes it from other entities, and isn't this way of existing exactly it's state? How can it not have a state of it's own?

{\em von} -- Well, by the principles of quantum mechanics, a state is represented by a ray in Hilbert space, and there is no such ray for the separate part, only for the two together.

{\em Eve} -- Well then maybe one hasn't gotten the principles right and one should rewrite them. After all these so called ``principles'' are human creations, aren't they? How about saying that the state of one part is that it is entangled with the other part in the given way, after all I'm related to my relatives in various ways, and that is part of {\em my\/} state. This way each photon, say, has as part of its state description it's entanglement with others.

{\em von} -- There's a problem with that for as I walk back and forth in this room, a distant photon polarization measuring experiment has either happened or not according to my instantaneous reference frame, and so this photon that just went trough this classroom had it's state description changed with my motion though I did not interact with it. That's not something you want of a state.

{\em Eve} -- But that happens with the two-photon state also, it's either entangled or not according to your motion.

{\em von} -- True, but the two-photon state is a non-local object, and there is nothing strange about it's description changing under change of reference. That's just different time-slice views of the same extended physical reality. For instance, a woman could change back and forth from happily married to a widow as she paces this way and that if her astronaut husband suffers a fatal accident on a far-away world. The state of marriage is not a local affair, it takes two, you know, who could be far apart.

{\em Eve} -- That's confusing a physical state with a legal one. One is married until legally declared a widow when the death or very probable demise of the husband is legally declared.

{\em von} -- You are mincing words Ms., Ms.?

{\em Eve} -- Adams Ma'am, Eve Adams.

{\em von} -- You must be from the Adams family\footnote{The dynasty of physics prophets, not the other one.} in which case you know what I mean. [{\em Why don't you {\em say} what you  mean then?}]  All physicists know how to calculate things, for instance, how, due to quantum non-locality,  the measurement of one part will change the state of the other one.

{\em Eve} -- But Professor you said just a little while ago that the other part doesn't have a state of it's own! How can you change something that doesn't exist in the first place?

{\em von} -- Ms. Adams, you're beginning to sound like my colleagues in the philosophy department. That's all  well and good but to do physics one doesn't need any of that. What is important is to be able to know how to calculate, make a prophesy, and check it experimentally.

{\em Eve} -- So you are telling me to shut up and calculate?

{\em von} -- Well, yes. Physics is all about prophesying  and verification, it {\sl describes\/} and does not explain.

{\em How about trying to {\em understand} what is {\em really} going on?}  Eve got fed-up with von Alteweib and with this kind of talk. She heard it all her life from her physicist relatives. She was tired of how physicists lie,\footnote{If you belive they don't lie, read \cite{cart}.} saying one thing, meaning another, and then saying everyone knew what they mean. [{\em Why not just {\em say\/} what you mean?}] Gathering up her notebook she rose from her chair and walked out of the classroom, getting quizzical looks from her classmates Diedre and Hugh. She just wanted to get away from it all, go to her little student's flat in La Huerta and rethink her career, maybe do something exciting, like joining the intelligence service. Getting into her miniature hybrid she drove off-campus and immediately got stuck in an enormous traffic jam. Sitting there she was getting more and more furious at the motorcyclists who were zipping through the traffic jam just like \ldots just like\ldots {\em why just like neutrinos zip trough light-years of lead. Yes, just like neutrinos! And if you shake nuclei, neutrinos pop out. Neutrinos are massless particles. If you shake atoms, photons come out, which are also massless. Photons hold atoms together. Well, EM interactions do, but you know what I mean! When you shake something then what holds it together should come out as massless particles. So neutrinos hold nuclei together, but wait, photons are spin one and bosons and neutrinos are spin half and fermions, two spins halves make spin one and two fermions make a boson, so a photon could actually be made of two neutrinos. Amazing! Brilliant! Everything is held together by neutrinos!\/} Suddenly she saw a  short-cut and a way out: she'll write a thesis so brilliant to be worth the Noble Prophet prize and get her Honorable Prophetess Declaration without interminable courses and exams, nor having to join some physics gang. She had to get home, but there was this traffic jam. Opening her window to see what was holding everyone back she surmised no apparent reason. Glancing sideways she saw a huge SUV next to her. The window of the SUV rolled down to reveal a striking woman of unfathomable age and an enigmatic smile. Eve noticed a small skull hanging from the rear-view mirror, but the woman seemed friendly enough.

{\em woman\/} -- Isn't it curious how the traffic can flow easily but the slightest disturbance at the right place will make it freeze and everyone gets stuck for a really long time.

{\em Eve\/} -- Well, yes, and just when I need to get home in a hurry to La Huerta to write a brilliant physics paper. My ideas are roiling in my head and I need to put them down!

{\em woman\/} -- As it happens I know these parts quite well and a few tricks to boot. There are strangely unnoticeable breaks in the traffic here and I can lead you through them to a point from which you can then go home quickly.

{\em Eve\/} -- Oh, if you do that I'll be eternally grateful to you! I'll be your slave!

The woman's smile broadened and Eve felt a sudden air of mysterious expectancy and a rare excitement.

{\em woman\/} -- Even better, follow me to my place which is closer. I have a truly great computer, I call it a q-circuit, on which you can start your work.

As Eve drove behind the SUV which by seeming magic could get around stopped vehicles in the jam, she felt an exhilaration and a joy that she would not feel again for\ldots

The notion of state in quantum mechanics is a triumph of practicality and conciseness, at a price. All physicist know how to use it in calculations, but there are conceptual quirks. Parts not having states of their own is one of these, and the status of mixed states is another. A qubit mixed state \(\frac 12 I\) can be decomposed into a mixture of pure states in infinitely many ways; there is an infinite-dimensional space of measures \(\mu \) on the set of rank-one projections \(P\) in \(\lC^2\) such that \(\frac12 I=\int P\,d\mu (P)\). Is  any one of these measures the {\em true\/} decomposition with the others being mathematical artifacts, or do they all have equal ontological status? One can also get this mixed state by partial trace on one of the parts of the singlet two-qubit state \(\frac1{\sqrt{2}}\left(\ket0\ket1-\ket1\ket0\right)\). Is this then the ``state" of one of the parts which it isn't supposed to have, or is this just a way of saying what a local observer ``sees" the state (which is supposed not to exist) to be? What about a state, of a photon prepared, say, by flipping a fair coin and preparing \(\ket 1\) if the result is heads, and \(\ket 0\) if not. Is this mixed state, after you've lost your notes as to which was created which way,  somehow different from that of one part of the two-photon singlet? There are bets on both sides in the literature. The coin-flip created one seems different. If one has a little daemon as a friend, one that can tell the polarization of a photon without measuring it, then he can tell you which way to turn your polarizer so that each photon of the coin created ensemble passes through, yet he would not be able (supposedly) to do so for the one part of the singlet since this photon is ``unpolarized" and does not have a polarization state of its own, as von Alteweib would want you to believe. Is this view legitimate? Do we have here another ontological quagmire that needs a little cleaning up?

The notion of a physical state is not needed to do physics. If the paradigm of an experiment is the preparations of a ``state" and then a subsequent measurement, all one needs to be able to predict is the joint probability of the preparation outcome and the measurement outcome. The ``state" is an interpolating agent that ``evolves" from the moment of preparation to the moment of measurement and is used in the calculus of joint probabilities. Such an agent need not have any real existence at all, and joint probabilities could in principle be calculated without invoking it. The usual quantum mechanical state is not just any interpolating agent,  it undergoes a deterministic unitary evolution and so the same state could have been created some time before its actual creation by different preparation, and can go on beyond the time of its destruction under measurement if this is not carried out. In other words the state is an autonomous being carrying no trace of its creation nor prescience of its demise.\footnote{Such states have peculiar properties, for instance in causal theories they must suffer instantaneous formal collapse under measurement.\cite{svet:FOP33.641}} Other kinds of state would have some imprint of the full story  of their existence.

One could object, legitimately, that the preparation, processing and measurement apparatus (the morphs of this paper) are physical objects having a ``state'' so one is still dealing with states. One has to assume a certain form of the Copenhagen  viewpoint and attribute a different ontology to these objects when banishing autonomous quantum states. This viewpoint was already expressed by Heisenberg.
 How to deal with the physics of apparatus if quantum states are banished is a separate issue which we won't take up here.

In the ``consistent histories'' approach to quantum mechanics of Murray Gell-Mann, James Hartle, Roland Omn\`{e}s and Robert B. Griffiths \cite{conhis}
 (see \cite{hall:ANYAS755.726} for a review) the notion of an interpolating quantum state is considerably weakened, though the formalism is based on the analysis of such a state via a {\em history\/} of temporally ordered measurements. Once the amplitudes of such histories are introduced, the intervening quantum states can be forgotten about and only the history amplitudes considered to be physically relevant. The initial and final states do still play a role, but these can be seen as just means to define sets of histories, which could be defined by some other means.

A benign version of the no physical state approach is the Heisenberg picture in ordinary quantum mechanics. True, there are state vectors, but they do not evolve in time. The observables evolve in time and the state vector is used as a  device to calculate joint probabilities, physically it is a sort of background. The background changes under measurement so the Heisenberg state vector still retains some characteristics that a physical state is usually conceived of as having, but some have been lost. This is another gauge-like change of mathematical description. We remind the reader that we're using the Heisenberg picture in most of the present discussion. The no-state view is of course here exemplified by the morph notation. Repeated indices replace lines and there is no ``evolution" of states between physical situations encoded by the morphs. In some of the recent works on the diagrammatic  treatment of quantum mechanics  much ado is found about the fact that lines linking nodes can be straightened, bent, or reshuffled in various ways, nodes flipped, and any number of other ``homotopy" transforms performed. Well, in the morph view, the are no lines, and so of course they can be  straightened, bent, or reshuffled in various ways {\em precisely because they are not there\/}! What part exactly of ``quantum mechanical essence" is then being formalized in this way?

The  \(I\) boxes of Alice and Diedre's qubit transfer scheme (\ref{AliceDiedre}) can be reexpressed by changing the linking line to it's opposite to get:
\begin{center}
\begin{picture}(130,110)(0,0)
\put(20,80){\framebox(30,20){\(I\)}}
\put(80,10){\framebox(30,20){\(I\)}}
\put(10,60){\line(1,1){20}}
\qbezier[30](40,80)(65,55)(90,30)
\put(100,30){\line(1,3){20}}
\put(20,70){\vector(1,1){0}}
\put(64,56){\vector(-1,1){0}}
\put(110,60){\vector(1,3){0}}\end{picture}
\end{center}
In Frobenious algebras the upper box is called an {\em evaluation\/} and the lower a {\em coevaluation\/}. Several authors have introduced Frobenious structures to express certain formal aspects of quantum mechanics. 
That the above diagram is equivalent to the direct line as in Alice's initial desire (Fig. \ref{Alice2Bob}) is know as the {\em zig-zag identity\/}. From the bare morph perspective (no lines) these objects are formalizing nothing at all. Formalizing nothing is not a bad idea, behold the numerical zero and the empty set (the decategorified zero), but one has to ask to what purpose one is doing this here.

Mathematics arose through abstraction from things of the world. Counting sheep, measuring land, and building structures were among the first practical uses of it. Now we've discovered that thing of the world, in their essence, behave quite differently, quantum mechanically, than how we thought before. Yet to describe this new behavior we're still acting as shepherds and pyramid builders. Our abstractions are still contaminated with the legacy of these old practices. Maybe it's time to begin anew and abstract directly from quantum behavior introducing as little as possible the structures that we've take so to heart as mathematicians. This text is a humble attempt in this direction. Cleansing out seems like a good thing to do.

Though on the one hand the no-physical-state attitude is possible,
it should also be possible, given a set of joint probabilities for preparation and measurement outcomes, to introduce {\em formal\/} interpolating states.\footnote{This is one of those statements for which I keep promising  myself to write up a true proof but never get around to doing it. In any case the subsequent discussion gives some indication of how it could be done.} Such formal states would not necessarily be of the autonomous type, and for some joint probabilities they {\em cannot\/} be of this type. Such states descriptions may necessarily incorporate a history of the state's existence.

Ivan is a mathematician and Kalysta a physicist; we think we know each one's ``state''.  Ivan meets Kalysta and says: ``I studied physics but then learned physicists think all series converge. Also their cavalier `proofs' look like math but are nothing like it at all. I was driven to desperation and  switched to mathematics.'' And Kalysta replies. ``I started as a mathematician but then  discovered physicists think all series converge and in fact the sum of the positive integers is minus a twelfth. This was infinitely more interesting than anything the mathematicians ever told me, so I decided to become a physicist.'' Well, now we know a lot more about Ivan and Kalysta and are better prepared to  understand what each may say or do as they meet.

At the other extreme of no-physical-state would be states so laden with information that even under entanglement  the whole {\em is\/} the sum of its parts. This can be done for a propic diagram of the physical type we're considering in this section. For simplicity consider such a diagram with no incoming or outgoing lines. This represents a complex  quantum amplitude calculated by means of (usual quantum) states and processes. For a start assume there is just one node with only outgoing lines and just one with only incoming lines.\footnote{If there are more than one of either type, by the associative property of props, they can be joined together into a single compound node.} We now assume each line represents a ``physical state'' whose description consists of the ``life history'' in a certain sense to be shortly described. When any number of these states enter a node, they share their histories, entering into a collective consciousness, so to speak, and leave the node with knowledge of the histories of the other participants, incorporating this into their own histories. In purely formal terms each line carries as its ``state'' the part of the diagram that can be reached by going  along the lines in the opposite, time-retrograde direction. Once these states enter the final node with only incoming lines, there is enough information available to construct the full diagram and so to calculate the amplitude. This way each line represents a state existing locally and impervious to what any other line carries at the same time and so ``entanglement'' is a superfluous notion. Since there is no entanglement of states and no ``collapse'' (there's ``growth'' instead as histories get accumulated) this scheme is safe from the type of objection von Alteweib raised to Eve's attempt to extend state description to include entanglement. Being truly local the description would change with reference frame only through a covariant change in the description of the nodes involved in the history, a frame dependence of the usual group-theoretic kind.  We can call this scheme the {\em local information explanation\/} (LIE) or the {\em personal story interpretations\/} (PSI) of physical states. Now should one believe the LIE or the PSI of \(\Psi \)? One could of course, it's mathematically consistent, but there's more than one price to pay.

The first price is ontological. It seems a tremendous burden for each particle, say a photon, to carry its whole history with it. Thinking slightly more deeply this doesn't even make sense, for a photon, as any other elementary particle, has no real identity; it can be absorbed, transformed, others can get emitted, so who is to carry the history of whom? This is even not bringing in the indistinguishability of identical particles. The neat propic diagrams are tremendous idealizations of real situations, it's hard to see how the LIE can be maintained in the real world. It may be possible but it doesn't seem worthwhile to  undertake the task of finding out.

There may however be some useful insights to be gained from the LIE. Consider the typical EPR experiment:

\begin{center}
\begin{picture}(140,90)
\put(55,0){\framebox(30,20){\(\Psi\)}}
\put(0,70){\framebox(30,20){\(\Lambda\)}}
\put(110,70){\framebox(30,20){\(\Omega\)}}
\put(65,20){\line(-1,1){50}}
\put(40,45){\vector(-1,1){0}}
\put(100,45){\vector(1,1){0}}
\put(75,20){\line(1,1){50}}
\end{picture}
\end{center}

For this to give the right complex number under LIE, the morphs labeled \(\Lambda\) and \(\Omega\) must act in concert which in the usual interpretations for states \(\Psi\) that violate Bell-type inequalities is attributed to a ``quantum mechanical non-locality of states". Now since under LIE the states are purely local the so called ``non-locality" must now be attributed to the ``measuring devices" that have \(\Lambda\) and \(\Omega\) respectfully as their corrersponding eigenstates. From the propic viewpoint one can consider the nodes \(\Lambda\) and \(\Omega\) as forming one node, using the associativity property and so look upon the above scheme as:

\begin{center}
\begin{picture}(190,110)
\put(80,0){\framebox(30,20){\(\Psi\)}}
\put(25,70){\framebox(30,20){\(\Lambda\)}}
\put(135,70){\framebox(30,20){\(\Omega\)}}
\put(90,20){\line(-1,1){50}}
\put(65,45){\vector(-1,1){0}}
\put(100,20){\line(1,1){50}}
\put(125,45){\vector(1,1){0}}
\qbezier(0,80)(0,110)(50,110)
\qbezier(0,80)(0,50)(50,50)
\qbezier(140,110)(190,110)(190,80)
\qbezier(140,50)(190,50)(190,80)
\put(50,50){\line(1,0){90}}
\put(50,110){\line(1,0){90}}
\put(95,100){\makebox(0,0){\(K\)}}
\end{picture}
\end{center}

The two nodes \(\Lambda\) and \(\Omega\) have been joined into a single node \(K\) as allowed by the associative property. This makes the EPR diagram equivalent to:
\begin{center}
\begin{picture}(30,110)
\put(0,0){\framebox(30,20){\(\Psi\)}}
\put(0,90){\framebox(30,20){\(K\)}}
\put(15,20){\line(0,1){70}}
\put(15,55){\vector(0,1){0}}
\end{picture}
\end{center}

This makes the shuac mathematician conjecture: \emph{Quantum mechanical non-locality is nothing but the associativity property of the quantum mechanical prop.} Non-locality can be paraphrased as ``separate thing act as one", which is also a paraphrase of  propic associativity.

\section{Into the Thick of it\protect\footnote{This section is loosely based on a talk I gave at Les Treilles in 2012 entitled ``Closed time-like curves in the presence of quantum mechanics", somewhat anti-\cite{deut:PRD44.3197}.}} \label{woods}

What Diedre discovered in a sudden classic quantum \emph{Aha!} moment is that any arrow carrying a Hilbert space \(\cH\) can be changed to \(\cB(\cH)\) and any node \(L\) can be changed to \(\lL=L\cdot \bar L\) and reinterpret the diagram in a consistent way. We call this process \emph{thickening}

\centerline{Thickening Diagrams}

\begin{center}
\begin{picture}(100,110)(0,0)
\put(0,40){\framebox(30,30)}
\put(15,55){\makebox(0,0){\(A\)}}
\put(15,0){\line(0,1){40}}
\put(15,70){\line(0,1){40}}
\put(25,20){\makebox(0,0){\(\cH\)}}
\put(25,90){\makebox(0,0){\(\cK\)}}
\put(15,25){\vector(0,1){0}}
\put(15,95){\vector(0,1){0}}
\put(50,40){\framebox(30,30)}
\put(65,55){\makebox(0,0){\(T_A\)}}
\put(65,0){\line(0,1){40}}
\put(65,70){\line(0,1){40}}
\put(80,20){\makebox(0,0){\(\cB(\cH)\)}}
\put(80,90){\makebox(0,0){\(\cB(\cK)\)}}
\put(65,25){\vector(0,1){0}}
\put(65,95){\vector(0,1){0}}
\end{picture}
\end{center}

\[T_A(M)=AMA^*\]

\centerline{Thickening Kets and Bras}

\begin{center}
\begin{picture}(90,50)(0,0)
\put(0,0){\framebox(30,20){\(\phi\)}}
\put(15,20){\line(0,1){30}}
\put(15,37){\vector(0,1){0}}
\put(50,0){\framebox(40,20){\(\ket\phi\bra\phi\)}}
\put(70,20){\line(0,1){30}}
\put(70,37){\vector(0,1){0}}
\end{picture}
\end{center}

\begin{center}
\begin{picture}(90,50)(0,0)
\put(0,30){\framebox(30,20){\(\phi\)}}
\put(15,0){\line(0,1){30}}
\put(15,17){\vector(0,1){0}}
\put(50,30){\framebox(40,20){\(\bra\phi\cdot\ket\phi\)}}
\put(70,0){\line(0,1){30}}
\put(70,17){\vector(0,1){0}}
\end{picture}
\end{center}

Any thickened propic diagram is a legitimate diagram expressing the same situation as the thin diagram. What thickening does is to now allow for mixed state processing. Instead of \(\psi\in \cH\) we have \(\rho\in \cB(\cH)\). The operators in the node need not be or the form \(T_A\), but can be any operator \(T\) from a tensor product of operator spaces (listed below) to another such. Thick diagrams are needed to handle open quantum systems and quantum channels. Thick diagrams embody another prop.\footnote{One can continue and thicken thick diagrams to form thick-thick diagrams. These process supeoperators. Diedre did not see any quantum mechanical need for these, though she might have been short-sighted. One could go on, thickening even more and more but what's the point?}   In thick diagrams arrow types, of which there are four now (\(\cB(\cH)\), \(\cB(\cH^*)\), \(\cB(\cH)^*\) and \(\cB(\cH^*)^*\)), and directions can be changed independently, as we now show. Remember we are dealing with finite dimensional Hilbert spaces.

Whereas for a Hilbert space the canonical isometry with its dual is antilinear, for the four spaces mentioned above
 we have in addition  \emph{linear} isometries:
\[\cB(\cH)\simeq \cB(\cH^*)\simeq \cB(\cH)^*\simeq \cB(\cH^*)^* \]
given by:
\[\ket i\bra j \leftrightarrow \bra \phi \mapsto \bracket \phi i \bra j\leftrightarrow A \mapsto \bra j A \ket i\leftrightarrow C\mapsto \bra jC'\ket i,\]
where \(C':\cH\to \cH\) is the \emph{dual} (not the adjoint) of \(C:\cH^*\to \cH^*\).

One can as always reverse a lines direction (and render it dotted) changing the associated space to it's dual. However, one can without reversing the line direction change the associated space to any one of the other three by placing appropriate linear isometries at the beginning and end of the line. As an example in
\begin{center}
\begin{picture}(200,70)
\put(0,10){\framebox(30,50){\(S\)}}
\put(170,10){\framebox(30,50){\(T\)}}
\put(30,30){\vector(1,0){70}}
\put(100,30){\line(1,0){70}}
\put(85,40){\(\cB(\cH^*)\)}
\end{picture}
\end{center}
we can change the line type to \(\cB(\cH)^*\) as follows:
\begin{center}
\begin{picture}(200,70)
\put(0,10){\framebox(30,50){\(S\)}}
\put(170,10){\framebox(30,50){\(T\)}}
\put(40,20){\framebox(20,20){\(L\)}}
\put(140,20){\framebox(20,20){\(R\)}}
\put(60,30){\vector(1,0){40}}
\put(100,30){\line(1,0){40}}
\put(30,30){\vector(1,0){10}}
\put(160,30){\vector(1,0){10}}
\put(85,40){\(\cB(\cH)^*\)}
\end{picture}
\end{center}
where \(L\) is the canonical linear isometry \(\cB(\cH^*)\to \cB(\cH)^*\) and \(R \) its inverse.

In what follows we shall not use this flexibility and all lines will correspond to \(\cB(\cH)\). Thick diagrams do not appear explicitly as such in the literature, though some diagrams that do appear really express thick notions.

Unbeknownst to Dutifully  Diligent Diedre, Ever Eavesdropping Eve had placed a undetectable quantum surveillance device following Diedre wherever she went. As soon as she saw that all arrows were equivalent in thick diagrams she immediately felt she had a grip on a possible time-travel scheme that she so longed for. Somewhere in her notebooks she remembered seeing the following thin diagram pretending to be thick: \footnote{Eve couldn't remember where she got the diagram, but it probably was based  on reference \cite{deut:PRD44.3197} by Deutsch, published in an obscure physics journal that charged money for access to its articles.}
\vfill\eject

\begin{figure}[h!]
\begin{center}
  \includegraphics[scale=0.2]{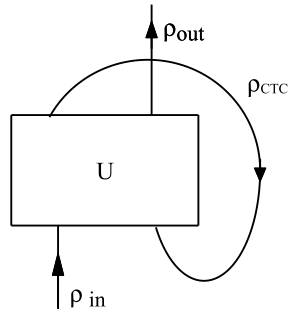}
\end{center}
\caption{Is this time-travel?}\label{thinD}
\end{figure}

She remembered the discussion she had with Baba Yaga about this diagram.
\vskip 3ex

\emph{Bab Yaga}  -- Poppycock!

\emph{Eve} -- Why poppycock, the math is perfect.

\emph{Bab Yaga} -- Your'e getting lost in math\footnote{Read \cite{lostinath}.}  just like Diedre. You can't have a unitary CTC since the universe is intrinsically an open quantum system.

\emph{Eve} -- What do you mean? Open quantum systems have environments. What's the environment for the universe?

\emph{Bab Yaga} -- No environment, you don't think of space time as embedded in a higher dimensional space to  explain its curvature. Curvature is intrinsic, and the ``openness" of the universe is intrinsic in the same way.

\emph{Eve} -- But string theorists do embed space-time in higher dimensional spaces.

\emph{Bab Yaga} --  You should not listen to the stringheads, there's not a shred of empirical evidence for their fantasy.

\emph{Eve} -- But it was said that Charlie was snared by a heterotic string, that's pretty empirical.

\emph{Bab Yaga} -- Where did you see that?

\emph{Eve} -- It was all over the physics blogs.

\emph{Bab Yaga} -- So now you're following the physics blogs! It's just hype and fake news! Charlie's sudden trip was another affair.

\emph{Eve} -- What other affair?

\emph{Bab Yaga} -- Explaining it would reveal sorcery secrets, and I can't do that.
\vskip 2ex
\emph{I'm tired of her sorcery secrets, but I \emph{must} find out what pulled Charlie to an event horizon, for there might be time-travel secrets therein. Still\ldots}
\vskip 2ex
\emph{Eve} -- Just what makes the universe an open quantum system?

\emph{Bab Yaga} -- There's an old fable about the origin of the universe, and expressed in modern jargon says that  space time is like a well prepared cappuccino from Piazza Sant'Eustachio in Rome, foamy that is. Within that foam there are tiny tiny tiny closed time-like curves (CTCs as physicists call them) and these make space-time an open quantum system.
\vskip 2ex
\emph{Aha! Tiny time machines at the basis of space-time.}
\vskip 2ex
\emph{Eve} -- But why do CTC do this?

\emph{Bab Yaga} -- It's just a fable, don't give it much thought, it could just as well have said that there are unicorns in space.
\vskip 2ex
\emph{I \emph{must} give it much thought.}

\vskip 3ex

As a thin diagram there is nothing special about Figure \ref{thinD}, it's just a partial trace of a unitary. What Deutsch does is to force this into being a time-travel diagram which depicts a state traveling back in time to interact with itself before returning to the present. The round-trip interpretation is reinforce by having the leftmost exit from \(U\) be the one that loops back to the rightmost entrance. This is purely formal. Why is this thin diagram not already a time-travel diagram? Well, the line of the partial trace does not \emph{carry} a quantum state, it's just a pictorial representation of the partial trace. Thin-diagrammatically the partial trace is given by:
\begin{center}
  \includegraphics[scale=0.2]{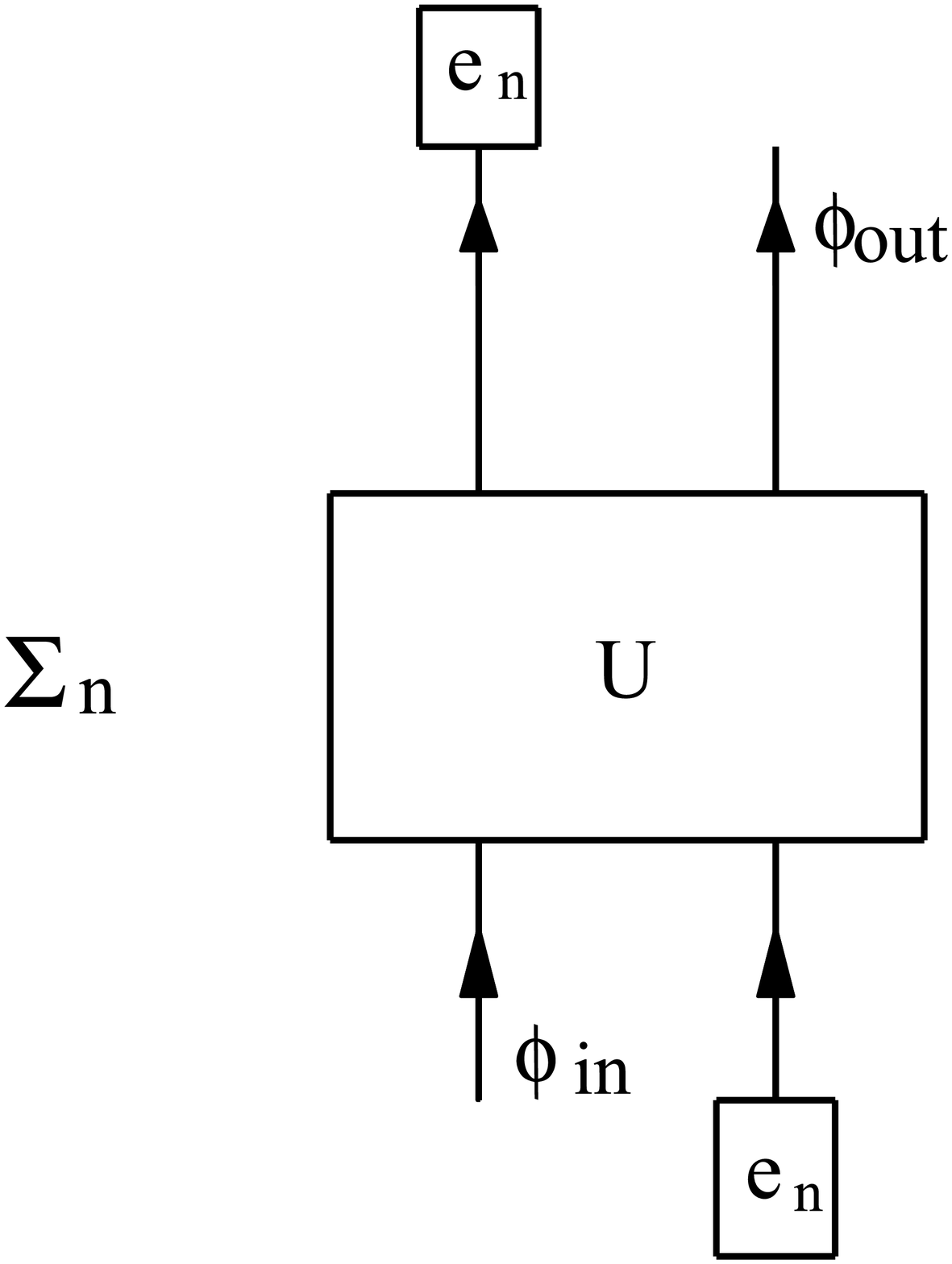}
\end{center}

\noindent where \(e_n\) is a qubit basis.

Thus there is no \(\rho_{ctc}\) nor any state that has any correspondence to the time-traveler's \(\phi_{in}\) and \(\phi_{out}\). Nevertheless, Deutsch forces the unitary node \(U\) to perform something for which quantum mechanics was  neither prepared nor adequate.  The node is just a theatrical prop hiding another construct. To insure at least mathematical consistency the following conditions are imposed.
\begin{eqnarray} \label{fixedp}
 \rho_{\rm ctc}=\Tr_2(U\rho_{\rm in}\otimes \rho_{\rm ctc} U^*),\\ \label{outstate}
 \rho_{\rm out}=\Tr_1(U\rho_{\rm in}\otimes \rho_{\rm ctc} U^*).
\end{eqnarray}

The first one, (\ref{fixedp}), expresses the idea that the state leaving the interacting region at the top to then go on a time-trip is the same that enters the region at the bottom. Time-travel does not change the traveler. The second one calculates what the round trip does to the traveler

Contemplating time-travel in a quantum world calls up many puzzles  beyond those just of time-travel itself. Some of these have never been addressed.
Consider the following version of Fig. \ref{thinD} where we explicitly show the world-line of the traveler. The interaction region has been rendered as a box with dotted sides. We have also introduced two measurement procedures \(M_1\) and \(M_2\) at the indicated points.
\begin{center}
  \includegraphics[scale=0.25]{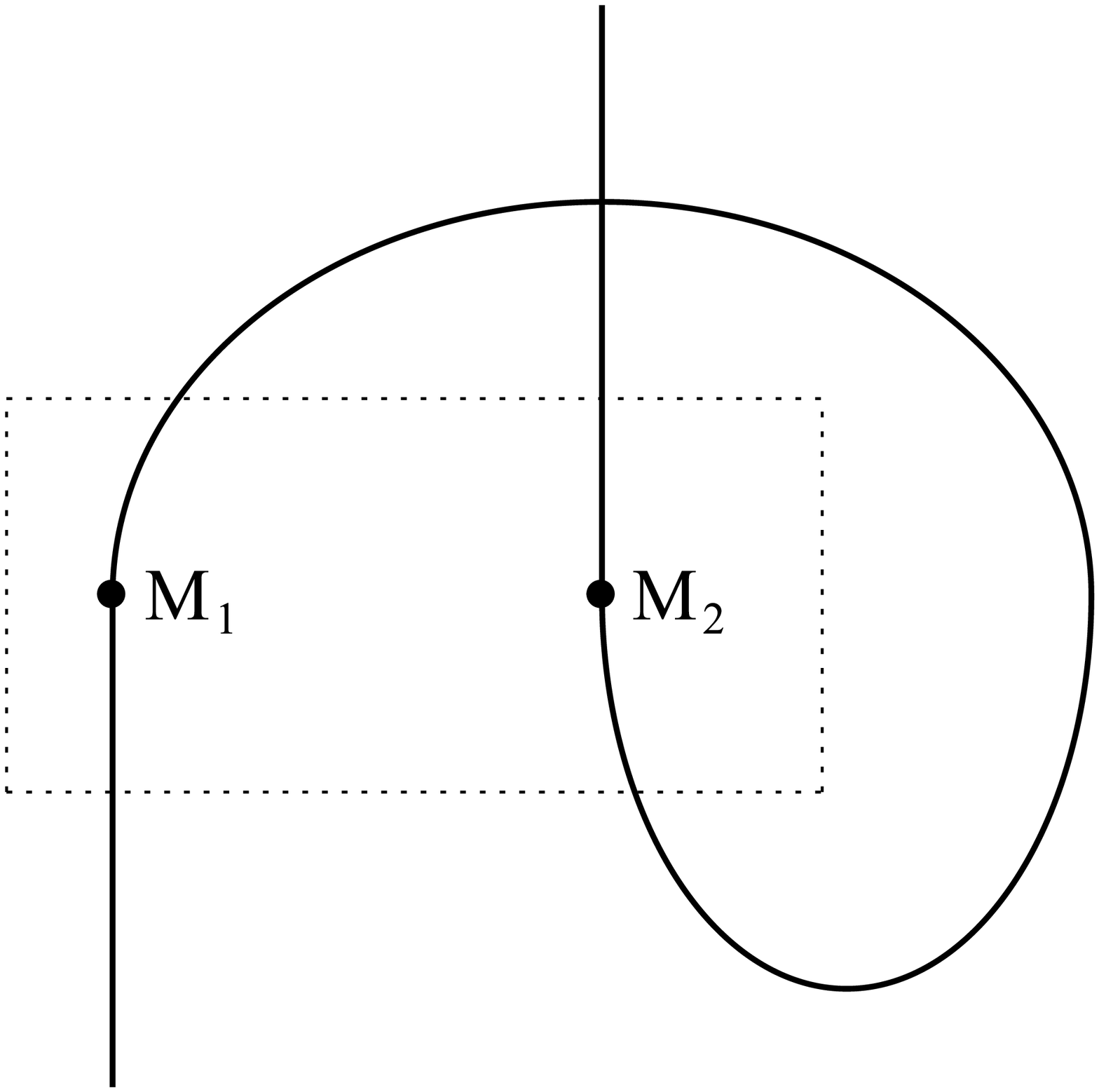}
\end{center}

Now all interactions are entangling for some states, thus even if the two selves that interact in the box are not entangled as they enter, they will generically get entangle inside and leave the box as such. So the question now is: \emph{What is the correlation between measurement results at \(M_1\) and \(M_2\)?}

On the one hand \(M_2\) is a measurement subsequent to \(M_1\) along the traveler's world-line, so there should be correlation of the same type as those of subsequent measurement along a chronology-preserving world-line.\footnote{More vividly, suppose the traveler at the \(M_1\) point gets shot and injured. He decides to go back in time and try to prevent this (Ignore the looming paradox.). He surely arrives at the \(M_2\) point still injured (presuming, as with Deutsch, that time travel doesn't change you). So obviously there have to be correlations between what happens at these points due to them being on a world-line of a physical system. } On the other hand the two measurements are simultaneous in a frame inside the box, so there should be correlations due to entanglement of the selves. So what is the final correlation in the end? Standard quantum mechanics has no answer to this, and neither do quantum time-travel schemes so far published in the literature.

The same conundrum occurs, though not as clearly,  with evolution. Presumably theere  is some kind of evolution along the world-line, but this has to also contend with the bipartite evolution within the box. Again standard, quantum mechanics offers no help.

Entanglement is superposition of coexistence. Generically, the two selves that leave the box are entangled, but further along the world line that leaves the box on the right, both the box and the time-traveling self are in the past light cone and so at that moment there cannot be any entanglement between the two selves\footnote{Assuming no further CTCs around.}. Moving through space-time with CTCs causes disentanglement.\footnote{Deutsch himslef has somewhat come to this conclusion. ``We must therefore reconsider that  the Universe as a whole may be described by a density operator  rather than a state vector." However, he does not consider this for universes without CTC's.}   Space-times with CTCs can be quite regular (viz. G\"odel's universe) so one could think there is no disentangling locally and that this would be a global effect. There is no immediately imaginable quantum scheme that can tackle this. One attitude could be: disentangling space-time is absurd so CTCs are impossible.
Another attitude could be: assume that in a quantum universe the same quantum laws apply whether there are CTCs or not and so assume that even without CTCs, space-time is an open quantum system. It is at least clear that allowing for CTCs, quantum mechanics has to be somehow modified or extended to treat the new situations.\footnote{One such proposal is \emph{Generalized Quantum Mechanics} \cite{hart:PRD49.6543} based on the consistent histories approach. We don't discuss it here as it falls somewhat outside our prop context.} Deutsch adopts a purely input-output approach. Though this is in keeping with a propic attitude, there are serious problems.

For those that love to get lost in math, there are no \emph{mathematical} difficulties in Deutsch's theory. For any \(\rho_{in}\) there always exists, by an appropriate fixed-point theorem, a solution \(\rho_{ctc}\) (not necessarily unique) to (\ref{fixedp}) after which one can calculate \(\rho_{out}\) by (\ref{outstate}) so you can predict in what state the time-traveler comes back. A major problem is that the transformation \(\rho_{\rm in}\mapsto \rho_{\rm out}\) is generically not implementable  by a linear quantum operator, which is again not the usual quantum behavior.\footnote{Non linear quantum theories generally allow for superluminal signals \cite{gisi:hpa62.363, polc:rpl66.397,svet:fop28.131}, but as one is seriously contemplating time-travel this objection loses most of its force.}  Another unusual feature assumed by Deutsch is that the density matrices \(\rho\) with labels shown in the diagram are not considered as mixtures of pure states, but as irreducible entities in their own right, the so called ``improper mixtures".\footnote{To quote from the article: "It is not intended to suggest a statistical ensemble"} To accept such gross distortions of quantum mechanics one needs a lot more than just consistency of the mathematics. This is another reason for the ``Poppycock" of the Russian sorceress.

Is there a way to talk about the quantum mechanics of time-travel without the problems of Deutsch's theory? The message of Deutsch's article seemingly is: with CTCs, quantum mechanics has to change and it has to change \emph{this way}. But is this the only way? One could take a hint from the idea that the universe is described by a density matrix and take this into account from the very beginning of setting up the input-output analysis.

It is precisely this that  Eve was wondering about. What if instead of unitaries one uses quantum channels and a thick version of the time-travel diagram, then maybe the poppycock can be overcome and a true time machine  built. So she drew the following:



\begin{center}
  \includegraphics[scale=0.3]{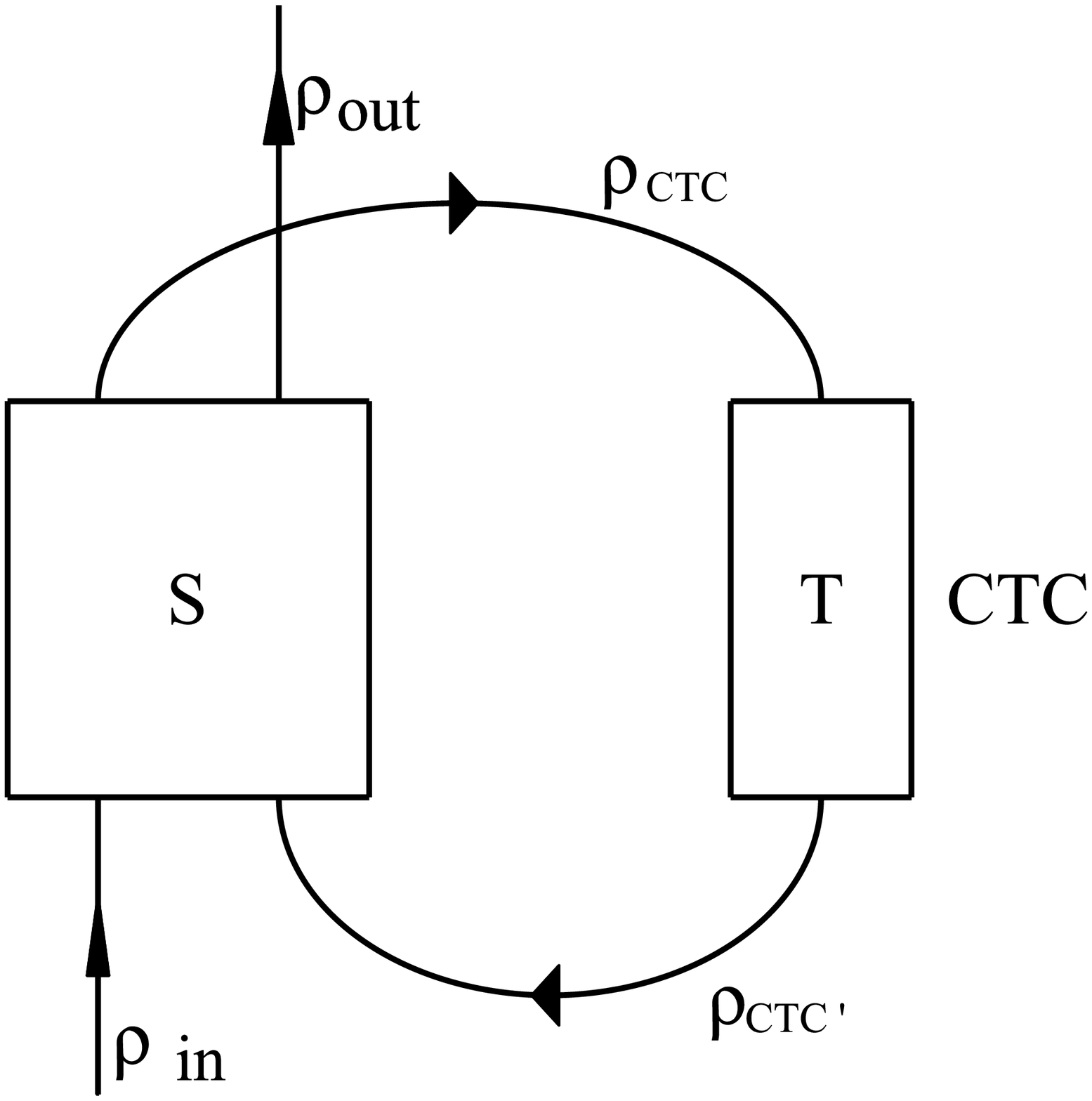}
\end{center}

Here both \(S\) and \(T\) are superoperators with \(T\) representing the time machine and \(S\), the \emph{direct channel}, the interaction of the time-traveler with him/her-self. We assume both are completely positive. There are now two states on the CTC, \(\rho_{CTC}\) the one that enters the time-machine and \(\rho_{CTC}{}'\) the one that leaves it in the past. time-travel itself will change your state. \emph{Prima facie} these states are no more real density matrices than in the Deutsch scheme, but one does have greater structural possibilities.

\emph{A-priori} one is basically at a loss to know which superoperators to pick. One can however get a hint from those thin diagrams where by post-selection on a measurement one effectively has time-travel\cite{svet:IJTP50.3903}.\footnote{See Czachor \cite{czac:PLA383.2704} for another approach} For qubits one such diagram is the following:

\begin{center}
\begin{picture}(90,150)(0,0)
\put(0,40){\framebox(30,30)}
\put(15,55){\makebox(0,0){\(U\)}}
\qbezier[40](60,20)(60,55)(60,90)
\put(10,10){\line(0,1){30}}
\put(10,27){\vector(0,1){0}}
\put(10,5){\makebox(0,0){\(\phi_{\rm in}\)}}
\put(10,70){\line(0,1){70}}
\put(10,147){\makebox(0,0){\(\phi_{\rm out}\)}}
\put(10,112){\vector(0,1){0}}
\put(30,90){\framebox(40,20){$\Psi_{xy}$}}
\put(30,115){\framebox(40,20){$\Psi_{xy}$}}
\put(30,0){\framebox(40,20)}
\put(85,115){\makebox(0,0){\({\cM}\)}}
\put(50,10){\makebox(0,0){\(\Psi _{00}\)}}
\put(20,70){\line(1,1){20}}
\put(20,40){\line(1,-1){20}}
\put(40,135){\line(0,1){15}}
\put(60,135){\line(0,1){15}}
\put(30,80){\vector(1,1){0}}
\put(30,30){\vector(-1,1){0}}
\put(60,55){\vector(0,-1){0}}
\put(60,145){\vector(0,1){0}}
\put(40,145){\vector(0,1){0}}
\end{picture}
\end{center}
Here \(\cM\) is a projective measurement on the four Bell states \(\Psi_{xy}\).  One line is dotted leading back in time to
simulate a time trip. With this, two of the depicted Bell states became channels. The upper channel shown corresponds (in relation to the computation basis and its dual) to the matrix
\(\sigma_{xy}/\sqrt{2}\) where \(\sigma_{xy}\) is an appropriate Pauli matrix. The lower channel corresponds to the matrix
\( I/\sqrt{2}\). When the measurement projects onto \(\Psi_{00}\) the overall time trip matrix is \(I/2\) and this is the case that is usually meant by ``post-selected" time-travel; the time-traveler is not changed upon arriving in the past. But as we are now considering that time-travel channels are not unitary, treating space-time as intrinsically open quantum systems, we no longer need to post-select on any one outcome, and any one can be considered as a case of post-selected time-travel with the time-traveler changed by the trip. As with the partial trace, and for analogous reasons, this in not ``real" time-travel but still just effective time-traavel. After a measurement has been made one can claim that time-travel has taken place, but there is no empirical way to prove or disprove this claim.
It  could be that real time-travel can happen spontaneously and we can only know of it after the fact being not able to empirically prove or disprove that it has happened.

When \(\cM\) projects onto \(\Psi_{xy}\) the (normalized) incoming state \(\phi\) is transformed into \(A_{xy}\phi\), where
\[A_{xy}=\frac1{\sqrt 2}\Tr_2((I\otimes \sigma_{xy})U)\]are  the Kraus operator of the channel defined by the measuring process. The probability of output \(A_{xy}\phi\) is or course \(\nu_{xy}=\|A_{xy}\phi\|^2\).

The thick diagram of this post-selective time-travel is now:

\begin{figure}[h!]
\begin{center}
  \includegraphics[scale=0.3]{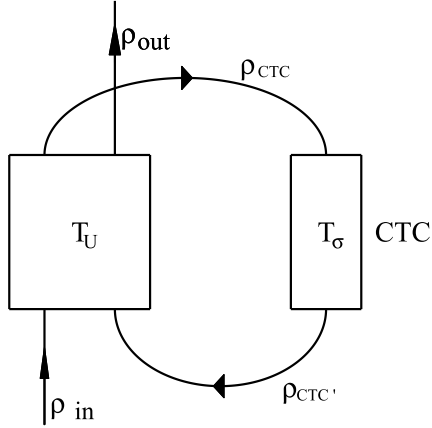}
\end{center}
\caption{Thick time-travel circuit}\label{thicks}
\end{figure}

The quantum channel  \(T_\sigma\) has Kraus operators that are multiples of \(\sigma_{xy}/\sqrt{2}\), explicitly
\[T_\sigma(M)=\sum_{x,y}\frac{\nu_{xy}}2\,\sigma_{xy}M\sigma_{xy}.\]
This is not a true rendition of the idea of a time machine, but can serve as a guide. One strange feature is that the inner workings of the time machine is dependent on what happens in the interactions of the two selves.\footnote{This circumstance has not been considered by science-fiction writers, as far as I know. Hint, hint.} To expedite the analysis we shall use ``\emph{future}" to refer to events further along the travelers journey, and ``\emph{past}" to those prior to the given subjective moment, be they in the future or past in relation to some standard chronology coordinate. One could envisage the time-traveler leaving the time machine at some space-time point to then continue under ordinary local dynamics to then in the \emph{future} interact with his \emph{past} self on another continent. That his \emph{past} self, after arriving at the future, would then be setting his time machine according to his \emph{future} and \emph{past} experiences in the past would seem unusual as one ordinarily thinks of time machines as construct not needing such adjustments.\footnote{In the absence of existing time machines, there is no real reason to think this unusual.} This is yet another type of non-locality though not necessarily of the quantum type.

To try to better understand some of this one can try fixed \(T_\sigma\) and the simplest is one with equal probabilities, that is:
\[T(M)=\sum_{x,y}\frac18\,\sigma_{xy}M\sigma_{xy}.\]

One has that for all density matrices \(\rho\)
\[T(\rho)=\frac12I.\]
and all states collapse to the maximally mixed one.

No information is sent to the past, but the past is influenced and the unitary channel transforms  into a non-unitary one.
 One is changed in a stochastic manner by interacting with one's \emph{past} and \emph{future} selves.  As simplistic as this channel may be it has some very desirable properties. In keeping within our thick philosophy we also replace the unitary channel \(T_U\) by an arbitrary completely positive channel \(S\).

\begin{enumerate}
\item \label{fixp} There's a Deutsch fixed point that is simply calculated.

The fixed point equation analogous to (\ref{fixedp}) is now:
\begin{equation}\label{tfixedp}
  \rho_{\rm ctc}=\Tr_2\left(S(\rho_{\rm in}\otimes T(\rho_{\rm ctc}))\right)=\Tr_2\left(S(\rho_{\rm in}\otimes I/2)\right),
\end{equation}
which simply \emph{computes} \(\rho_{\rm ctc}\).

And now
\begin{equation}\label{rhout}
  \rho_{\rm out}=\Tr_1\left(S(\rho_{\rm in}\otimes T(\rho_{\rm ctc}))\right)=\Tr_1\left(S(\rho_{\rm in}\otimes I/2)\right).
\end{equation}

Note that one need not compute \(\rho_{\rm ctc}\) to compute \(\rho_{\rm out}\) since whatever  \(\rho_{\rm ctc}\) may be, \(\rho_{\rm ctc'}\), which, along with \(\rho_{\rm in}\), is what determines \(\rho_{\rm out}\), is always the completely mixed state \(I/2\).

From this we also conclude:
\item \label{rrlin}\(\rho _{\hbox{in}}\mapsto \rho _{\hbox{out}}\) is linearly implementable.

This follows immediately from (\ref{rhout}).

\item \label{prmx}There's no need of improper mixtures.
\end{enumerate}

When Eve discovered this she was at once exited and disgruntled. Exited in seeing that there \emph{was} a time-travel process that respected standard quantum mechanics and disgruntled in that upon entering this time machine she could come out in the past as anything at all as her \emph{future} self. She could come out as a fire breathing six-headed snarling  sea serpent, or even\ldots

``Oh no!" she cried ``I could come back as Baba Yaga. Maybe in the \emph{future} I will time-travel and come back as Baba Yaga. No wonder we got stuck together. No that can't be, she is much wiser than I am, so from where would I have gotten that wisdom  to then pass it on to myself? Why from \emph{myself} of course, she (my \emph{future} self) has been teaching me (my present self) all sorts of things. But then this would be wisdom \emph{sui generis} created by nobody. No, it can't be true that Bab Yaga is my \emph{future} self!"

Having side stepped (or so she thought) this seeming paradox she relaxed and went on with her research wishing she could confide in Diedre to use her greater mathematical skills and missing her erstwhile colleagues  when they studied quantum mechanics (or what passed for it) together..

A time-travel channel that satisfies the above listed properties (\ref{fixp}-\ref{prmx}) when coupled to any channel \(S\) we will call a \emph{universal channel}.  We have:

\begin{enumerate}
\item Any channel that collapses to a point is a universal channel.

This is clear from equations (\ref{tfixedp}) and (\(\ref{rhout})\) by simply replacing \(I/2\) by the fixed state.

\item Any channel composed with a universal one is universal.

Indeed let \(T=T_1\circ T_2\)  where one is universal, then by propic associativity we can incorporate the other one into \(S\) to form a new direct channel.

\item \label{limu} A limit of universal channels is universal.
 \end{enumerate}

 When one fine evening Eve discovered the above facts she got animated again. ``I can build a time machine that turns everything into \emph{me}! That way I'd go back unchanged." But technically  this seemed doubtful and daunting. ``Wait! I can build a trillion much simpler time-machines\footnote{See the chapter ``Table-top time-travel" in the much awaited book \emph{Quantum Quiddity} by Alice, Bob, Charlie, Diedre and Eve, published by Yaga Press Ltd. to come out sometime early in the past decade.} each one transforming anything into a microscopic part of me. This way I'd get disassembled and sent back in pieces and get reassembled again, much like those ridiculous ``beam me up"  schemes in the \emph{Star Wreck} movies and series." But then she got just as crestfallen as before, for this also seemed to be just as daunting. ``I give up," she cried,
only by \emph{magic} can this be done! That's how Baba Yaga would do it." In the depth of her despair she suddenly felt an empowering thought.  ``Time to act! I'm tired of being passive and just spying on my former friends. The world is not made of \emph{beables} but of \emph{doables}, and Aristotle, bambino,  when you said \emph{All that is, is} you vacillated,  and should have said \emph{All that is, does!}  \emph{Sum ergo facio! } May the Quantum be with me!" The story \ldots

 One can use item (\ref{limu}) above to seek universal channels. Let \(T: \cB(\cH)\to\cB(\cH)\) be a completely positive map. There are positive integers \(n_k\to \infty\) such that \(T^{n_k}\to K\) a unique completely positive idempotent. One can look for universal channels among the idempotents. For a qubit there are three types of completely positive idempotents \(J\):

 \begin{enumerate}
\item Those that collapse to a point. \(J(A)=\Tr(A)\rho_0\)
\item Projective measurement: \(J(A)=P_1AP_1+P_2AP_2\) where \(P_1,P_2\) is a resolution of the identity.
\item The identity \(I\).
\end{enumerate}

Of these only the first one is universal. This does not mean that any \(T\) whose limiting idempotent is of the first type is universal and most likely none such are that are not already the idempotent. The situation in higher dimension has not been explored. It seems though that the possibilities for quantum
linearity preserving time-travel circuits is disappointedly limited.

In all fairness, not all Deutsch circuits lead to nonlinearity. Consider the control-\(V\)  unitary  with trace on the control qubit.

\begin{center}
\begin{picture}(70,80)(0,0)
\put(45,30){\framebox(20,20)}
\put(55,40){\makebox(0,0){\(V\)}}
\put(55,0){\line(0,1){30}}
\put(55,50){\line(0,1){30}}
\put(15,40){\circle{25}}
\put(27.5,40){\line(1,0){17.5}}
\put(27.5,40){\circle*{4}}
\end{picture}
\end{center}
As a Deutsch circluit, any density matrix \(\lambda_1P_0+\lambda_1P_1\), diagonal in the computation basis, is a Deutsch fixed point. One has:
\[\rho_{out}=\lambda_0\rho_{in}+\lambda_1V\rho_{in}V^*\]

Thus there are Deutsch circuits that do not call for deformed quantum mechanics, however this would not be a satisfactory answer to what quantum mechanics would be in the presence of CTCs. It would mean that any time traveler has to interact with himself in very specific ways so as not to spoil quantum mechanics.\footnote{This is another hint for would-be science fiction writers.}

In  the linearity preserving situations above, be they Deutsch or not, the time travel channel results in a fixed density matrix entering the interaction region in the past, independently of what \(\rho_{in}\) is. In a sense no information is sent to the past in a particular temporal round trip. This does n;ot mean that one cannot send messages to the past. All one needs to do is to set up two time machines each one sending a different fixed state to the past, and by alternating the use of the two one can send a binary message. Paradoxes with being able to communicate to the past still remain.

What we've explored in this section is not really time-travel but only the props of time-travel. Props in two senses, as a mathematical scaffold by which to make some calculations (the shuac approach) and as theatrical props, just cut-outs that give the appearance of something different but in end are just props. If CTCs have any real role in physics, what we've done here probably has very little to do with the real stuff. We've  learned  nothing about the possibility of time-travel but a lot about its props. Pretty much the same goes for quantum mechanics as a whole. and the story ends here.
\vskip 2ex
But wait! Something else is going on\ldots
\vskip 3ex

Eve is hacking into Baba-Yaga's q-circuit computer, trying one evil sounding password after another and, after spending hours and in pure exhaustion, just hits the keys randomly (\emph{why does she still use keys instead of telepathy?}) which suddenly resulted in what seemed to say ``leg'o'bone" in Cyrillic. She is in. \emph{Very  suspicious, no sophisticated q-tricks against intruders.} There was one non-encrypted file labeled ``Charlie's wobbly time-warp wanderings" Inside she   finds a hideous plan to send Charlie to an event horizon as soon as he approached the L4 Earth-Moon colony on a tourist trip. Further on, she reads ``\ldots  Charlie got too close to time-travel, though he never \ldots have to send him away\ldots" There was a lot of runic gibberish part of which Eve recognizes as a quantum magic spell to ``ravel" (or maybe ``unravel") time.  She quickly memorizes it and sneaks out of Bab-Yaga's lair. \emph{That was too easy!} Bravely  ignoring this very  disturbing thought she decides to go on.

Taking the cash she earned moonlighting as a PI unbeknownst to Baba Yaga (\emph{Why did she never find this out about me?}) she books a passage on a space tourism flight heading toward the  L4 point following Charlie's path. Getting close she  goes on a supervised apace-walk  and uses the quantum magic spell to open up a wormhole adding the extras needed to bring Charlie back. She suddenly finds herself  heading toward the   wormhole with what seems like a huge rotating cosmic string on the other side, hoping against all odds to finally free herself from slavery. Passing out from the strong magnetic fields cursing through her brain she almost remembers a long forgotten day...

\centerline{\(\vdots\)}

{\em Eve} -- So you are telling me to shut up and calculate?

{\em von} -- Well, yes. Physics is all about prophesying  and verification, it {\sl describes\/} and does not explain.

{\em How about trying to {\em understand} what is {\em really} going on?}  Eve was fed-up with von Alteweib and with this kind of talk. She heard it all her life from her physicist relatives. She was tired of how physicists lie, saying one thing, meaning another, and then saying everyone knew what they mean. [{\em Why not just {\em say\/} what you mean?}] Gathering up her notebook she rises from her chair and is about to walk out but something holds her back. ``I'll show that von Alteweib where  crawfish spend their winter."\footnote{She couldn't figure out where she got that expression, but some vague image of a Russian \emph{(why Russian?)} dominatrix came to mind.} So she sits through the lecture feeling happy  for no apparent reason and even feels some warmth toward Diedre and her mathematical fluff. She is thinking up a multitude of arguments to put the whole confusing quantum confusion straight, such as: ``quantumness is in space-time and not in the particles". Leaving the campus she walks to her miniature hybrid when she sees a huge SUV driven by a striking woman who gives her a warm but enigmatic smile. Heartened by the smile she feels the presence of some deep wisdom held up within her. She almost runs over Charlie crossing the street, stops to pick him up, and they start on their plan to show von Alteweib what's really what in the quantum world.

\section{And On and On}
The story does not end here \ldots


\newpage

\section*{Acknowledgements}This research received partial financial support by the Conselho Nacional de Desenvolvimento Cient\'{\i}fico e Tecnol\'ogico (CNPq), and the Funda\c{c}\~ao de Amparo \`a Pesquisa do Estado do Rio de Janeiro  (FAPERJ).

\end{document}